\documentclass[twocolumn,floatfix,amsmath,amssymb,superscriptaddress]{revtex4}
\usepackage[dvips]{graphicx}
\usepackage{amsfonts}
\usepackage{dcolumn}
\usepackage{bm}
\usepackage{color}
\usepackage{epsfig}
\usepackage{hyperref}
\usepackage{bbold}

\newcommand{\be}{\begin{equation}}
\newcommand{\ee}{\end{equation}}
\newcommand{\bea}{\begin{eqnarray}}
\newcommand{\eea}{\end{eqnarray}}

\def\G{\Gamma}
\def\d{\delta}
\def\D{\Delta}
\def\e{\epsilon}

\def\th{\theta}
\def\k{\kappa}

\def\m{\mu}
\def\n{\nu}

\def\p{\pi}

\def\r{\rho}
\def\s{\sigma}
\def\S{\Sigma}

\def\vf{\varphi}
\def\F{\Phi}

\def\w{\omega}
\def\W{\Omega}
\def\q{\psi}
\def\Q{\Psi}

\def\calgS{\mbox{$\mathit{\Sigma}$}}

\def\bld{{\mathbf d}}

\def\blp{{\mathbf p}}

\def\blr{{\mathbf r}}

\def\blE{{\mathbf E}}

\def\callE{\mbox{$\mathcal{E}$}}

\def\callG{\mbox{$\mathcal{G}$}}

\def\callI{\mbox{$\mathcal{I}$}}
\def\callJ{\mbox{$\mathcal{J}$}}

\def\callT{\mbox{$\mathcal{T}$}}

\def\ua{\uparrow}
\def\da{\downarrow}

\def\de{\partial}

\def\bra{\langle}
\def\ket{\rangle}

\def\1op{\hat{\mathbbm{1}}}
\def\nn{\nonumber}

\begin{document}

\title{Real-time dynamics of Auger wavepackets and 
decays in ultrafast charge migration processes}

\author{F. Covito}
\affiliation{Max Planck Institute for the Structure and Dynamics of 
Matter and Center for Free-Electron Laser  Science, Luruper Chaussee 
149, 22761 Hamburg, Germany}  
\author{E. Perfetto}
\affiliation{CNR-ISM, Division of Ultrafast Processes in Materials 
(FLASHit), Area della ricerca di Roma 1, Monterotondo Scalo, Italy}
\affiliation{Dipartimento di Fisica, 
Universit\`a di Roma Tor Vergata, Via della Ricerca Scientifica, 00133 Rome, Italy}
\author{A. Rubio}
\affiliation{Max Planck Institute for the Structure and Dynamics of 
Matter and Center for Free-Electron Laser  Science, Luruper Chaussee 
149, 22761 Hamburg, Germany}  
\affiliation{Center for Computational Quantum Physics (CCQ), The 
Flatiron Institute, 162 Fifth avenue, New York NY 10010} 
\affiliation{Nano-Bio Spectroscopy Group, Universidad del Pa\'is Vasco, 
20018 San Sebastian, Spain}  
\author{G. Stefanucci}
\affiliation{Dipartimento di Fisica, Universit\`a di Roma Tor Vergata, Via della Ricerca Scientifica, 00133 Rome, Italy}
\affiliation{INFN, Sezione di Roma Tor Vergata, Via della Ricerca Scientifica 1, 00133 Roma, Italy}
\date{\today}

\begin{abstract}
The Auger decay is a relevant recombination channel during the first few 
femtoseconds of molecular targets impinged by attosecond XUV or soft 
X-ray pulses. 
Including this mechanism in
time--dependent simulations of  charge--migration 
processes
is a difficult task, and  
Auger scatterings are often ignored altogether. In this work we 
present an advance of the current state-of-the-art by putting forward a 
real--time approach based on nonequilibrium Green's functions
suitable for first-principles  calculations of 
molecules with tens of active electrons. 
To demonstrate the accuracy of the method we report comparisons 
against accurate grid simulations 
of one-dimensional systems.
We also predict a highly asymmetric profile of the  
Auger wavepacket, with a long tail exhibiting ripples
temporally spaced by the inverse of the Auger energy.

\end{abstract}

\maketitle


The sub-femtosecond dynamics of the hole density created by an 
ionizing attosecond XUV or soft X-ray pulse 
precedes any nuclear rearrangement and dictates the 
relaxation pathways of the underlying molecular structure~\cite{Calegari336,uiberacker2007attosecond}. 
This ultrafast charge oscillation, 
also referred to as ultrafast charge migration (UCM), 
is driven exclusively by electronic correlations
up to a few 
femtoseconds~\cite{KuleffCederbaum,PhysRevLett.117.093002,BreidbachCederbaum,PhysRevA.91.051401,Nagaya.2010}.
At these time scales  the Auger scattering is the only possible 
energy--dissipation mechanism and, in addition to shake-up and polarization 
effects~\cite{PhysRevA.66.042715},  a relevant recombination channel.

Recent advances in pump-probe spectroscopy made possible to follow 
the Auger decay in atomic 
targets~\cite{Uphues2008,drescher2002time,uiberacker2007attosecond,zherebtsov2011attosecond,Schins-PhysRevLett.73.2180}.  
Accurate measurements have been performed and successfully 
interpreted in terms of transitions between excited cationic states. 
The theory behind these
experiments shows that the Auger electron is a ``courier'' of 
the complex dynamics occurring in the 
parent cation~\cite{PhysRevLett.91.253001,Kazansky.2009,Kazansky.2011}. 
Unfortunately, {\em ab~initio} analysis relying on 
many-electron eigenfunctions and eigenvalues 
are possible for single atoms but
become soon prohibitive for larger systems.
In fact, first-principles 
approaches that include Auger
scatterings in the UCM dynamics of 
molecules have not yet been developed.

Time--Dependent Density Functional 
Theory~\cite{RungeGross:84,Ullrich:12,Maitra.2016} (TDDFT) is the method of choice for 
large scale simulations. However, the vast majority of TDDFT calculations are 
performed using an adiabatic exchange-correlation (xc) potential, i.e., a 
functional of the instantaneous density. As shown in 
Ref.~\cite{PhysRevB.86.045114}, adiabatic 
approximations are unable to capture the Auger 
effect~\cite{auger-tddft}.
Learning how to include memory effects in the xc 
functional is a major 
line of research to which the present work could provide  
new insights.   

In this Letter we lay down a first-principles real-time 
NonEquilibrium Green's Function~\cite{kadanoff1962quantum,svl-book} 
(NEGF) approach 
which incorporates Auger scatterings in the 
UCM dynamics of molecules hit by attosecond pulses. 
In analogy with the NEGF formulation of quantum transport where the 
dynamics of electrons in the junction is simulated without dealing 
explicitly with the electrons in the 
leads~\cite{mssvl.2009,mssvl.2008,LPUvLS.2014}, we close the NEGF 
equations on the molecule and deal only partially with the degrees of 
freedom of the Auger electrons. 
The computational effort 
changes slightly with respect to previous NEGF 
implementations~\cite{PUvLS.2015,C60paper2018,PSMP.2018}, thereby 
making possible to simulate the UCM of
molecules with tens of active electrons.

We demonstrate that the approach
well captures qualitative and quantitative aspects of the Auger 
physics through comparisons against real-time simulations of 
one-dimensional (1D) atoms on a grid.
The Auger wavepacket can, in principle, be reconstructed from NEGF through a 
postprocessing procedure. For 3D molecules such procedure is 
numerically (too) demanding but for the considered 1D atom the 
calculation is doable 
and the agreement with the full-grid results is again satisfactory.
Interestingly, we highlight a universal feature of the  
asymmetric Auger wavepacket, namely 
a long tail with superimposed ripples temporally spaced by the inverse 
of the Auger energy.

\begin{figure}[t]
\centerline{
\includegraphics[width=0.49\textwidth]{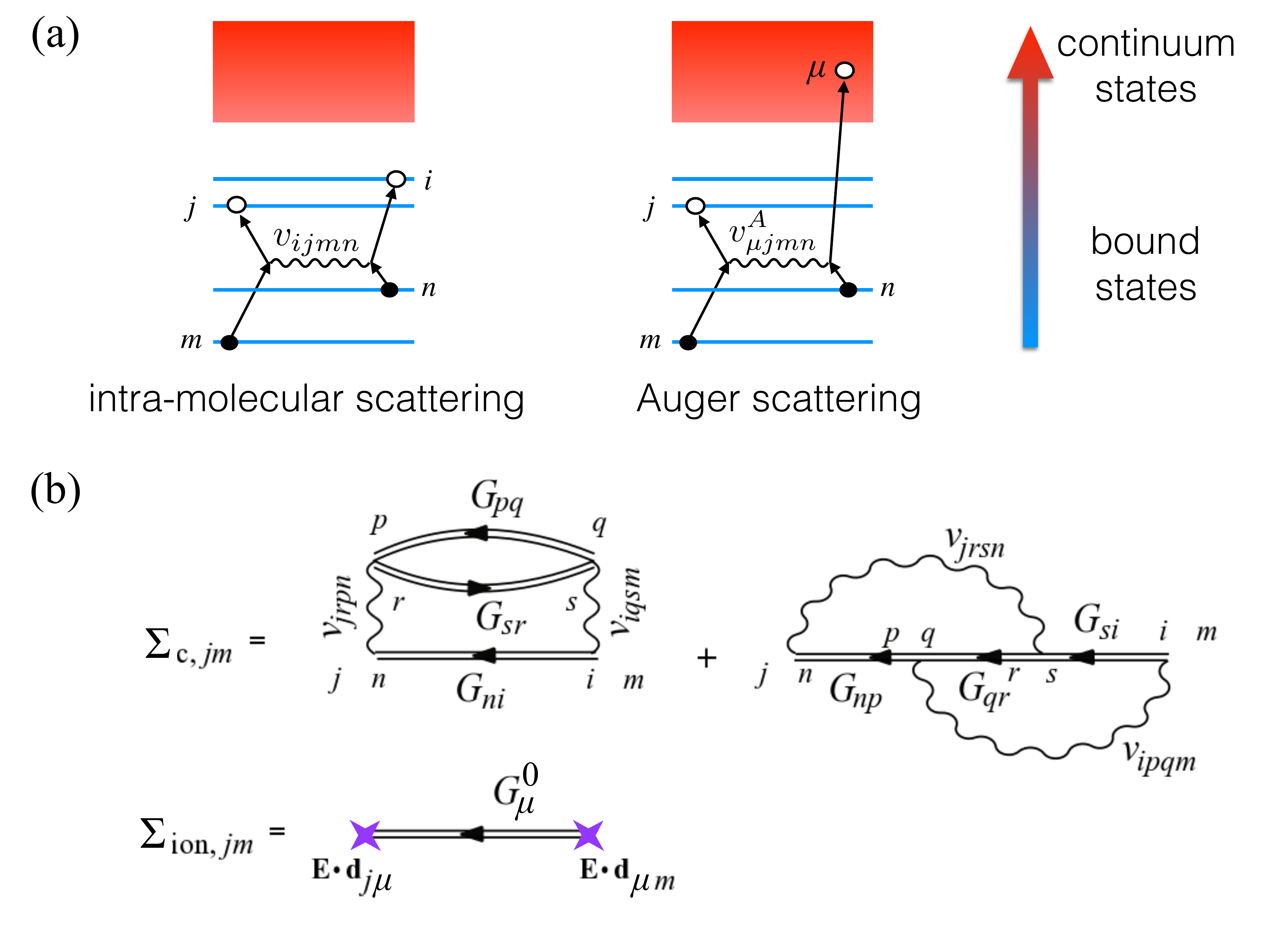}}
\caption{(a) Schematic illustration of intra-molecular (left) and Auger (right) 
scattering. (b) Correlation self-energy in the 2B approximation (top) 
and ionization self-energy (bottom).}
\label{scatt+sigma}
\end{figure}

{\em Method}:
We consider a finite system (an atom or molecule) with 
single-particle Hartree-Fock (HF) basis $\vf_{i}(\blr)$ for 
bound electrons and $\vf_{\m}(\blr)$ for electrons in the continuum 
(for simplicity we work with spin-degenerate systems).
Let $\hat{c}_{i\s}$ ($\hat{c}_{\m\s}$) be the 
annihilation operator for an electron on $\vf_{i}$ ($\vf_{\m}$) with spin $\s$.
In the absence of external fields the total Hamiltonian 
\be
\hat{H}^{\rm eq}=\hat{H}_{\rm bound}+\hat{H}_{\rm Auger}+\hat{H}_{\rm 
cont}
\label{eqham}
\ee
is the sum 
of the bound--electrons  Hamiltonian $\hat{H}_{\rm bound}=
\sum_{\substack{ij\\ \s}}h_{ij}\hat{c}^{\dag}_{i\s}\hat{c}_{j\s}
+\frac{1}{2}\sum_{\substack{ ijmn\\ \s\s'}}
v_{ijmn}$ $\hat{c}^{\dag}_{i\s}\hat{c}^{\dag}_{j\s'}\hat{c}_{m\s'}\hat{c}_{n\s}$, the Auger 
interaction $\hat{H}_{\rm Auger}=\sum_{\substack{ ijm\m\\ \s\s'}}
v^{A}_{ijm\m}
\left(\hat{c}^{\dag}_{i\s}\hat{c}^{\dag}_{j\s'}\hat{c}_{m\s'}\hat{c}_{\m\s}
+{\rm h.c.}\right)$ and a free-continuum part $\hat{H}_{\rm 
cont}=\sum_{\m\s}\e_{\m}\hat{c}^{\dag}_{\m\s}\hat{c}_{\m\s}$.
Here $h_{ij}$ are the one-electron integrals,  $\e_{\m}$ are the 
continuum single-particle energies and
$v_{ijmn}$ ($v^{A}_{ijm\m}$) are the four-index Coulomb integrals 
responsible for intra-molecular (Auger) scatterings, see 
Fig.~\ref{scatt+sigma}(a). 

The system is perturbed either by the sudden removal of a bound 
electron or by an external laser field. In the dipole approximation the 
laser--system interaction reads 
\be
\hat{H}^{\blE}(t)=\hat{H}^{\blE}_{\rm bound}(t)+
\hat{H}^{\blE}_{\rm ion}(t),
\label{laserham}
\ee
where $\hat{H}^{\blE}_{\rm bound}(t)=\blE(t)\cdot
\sum_{\substack{ij\\ \s}}\bld_{ij}\hat{c}^{\dag}_{i\s}\hat{c}_{j\s}$ describes 
intra-molecular transitions whereas  
$\hat{H}^{\blE}_{\rm ion}(t)=\blE(t)\cdot
\sum_{\substack{i\m\\ 
\s}}\left(\bld_{i\m}\hat{c}^{\dag}_{i\s}\hat{c}_{\m\s}+{\rm 
h.c.}\right)$ is responsible for ionization. The vector 
$\bld_{ij}$  ($\bld_{i\m}$) is the matrix element of the dipole 
operator between states $\vf_{i}$ and $\vf_{j}$ ($\vf_{\m}$). In 
Eqs.~(\ref{eqham}) and (\ref{laserham}) we are discarding the 
off-diagonal elements $h_{i\m}$, $h_{\m\m'}$ and 
$\bld_{\m\m'}$ as well as all Coulomb integrals with two or more 
indices in the continuum. We anticipate that this  
simplification affects only marginally the results presented below.

The electron dynamics is simulated using NEGF.
Without Auger scatterings the equation of 
motion for the one--particle density matrix 
$\r_{ij}(t)=\bra\hat{c}^{\dag}_{j\s}(t)\hat{c}_{i\s}(t)\ket$ (with indices 
in the bound sector) has been 
derived elsewhere~\cite{PUvLS.2015} and reads
$\dot{\r}=-i\left[h_{\rm HF}[\r],\r\right]
-\callI[\r]-\callI^{\dag}[\r]$. 
Here the HF Hamiltonian $h_{{\rm HF}}(t)\equiv 
h+V_{{\rm 
HF}}(t)+\blE(t)\cdot\bld$ is a functional of $\r$ through the HF 
potential $V_{{\rm HF},ij}(t)=\sum_{mn}\r_{nm}(t)w_{imnj}$, with 
$w_{imnj}\equiv 2v_{imnj}-v_{imjn}$. Dynamical correlation and 
ionization processes are described by the generalized collision 
integral 
\be
\callI(t)=
\int_{0}^{t}\! d\bar{t}\!\left[\S^{>}(t,\bar{t})G^{<}(\bar{t},t)-
\S^{<}(t,\bar{t})G^{>}(\bar{t},t)\right],
\label{collwoAuger}
\ee
where $\S^{\lessgtr}\equiv\S^{\lessgtr}_{\rm c}+\S^{\lessgtr}_{\rm ion}$ is the sum of the 
lesser/greater correlation ($\S_{\rm c}$) and ionization ($\S_{\rm ion}$) 
self-energies. Both are time-nonlocal functionals of $\r$ through the Generalized 
Kadanoff-Baym Ansatz~\cite{PhysRevB.34.6933} (GKBA), see 
Appendix~\ref{appendixA} for details. 
Figure~\ref{scatt+sigma}(b) illustrates the 
diagrammatic representation of $\S_{\rm c}$ in the second--Born (2B) 
approximation and $\S_{\rm ion}$. The computational cost of 
these NEGF calculations scales like $N_{t}^{2}N_{\rm bound}^{\mathfrak{p}}$ where 
$N_{t}$ is the  number of 
time-steps,  $N_{\rm bound}$ the number of HF bound states 
and the power $3\leq \mathfrak{p}\leq 5$ depends on how sparse
$v_{ijmn}$ is. Real--time simulations of, e.g., organic 
or biologically relevant molecules
can easily be carried out up to $30\div 
40$ femtoseconds~\cite{PSMP.2018}.

The inclusion of Auger scattering processes leads to 
a coupling between the density matrix $\r(t)$ and the occupations 
$f_{\m}(t)=\bra\hat{c}^{\dag}_{\m\s}(t)\hat{c}_{\m\s}(t)\ket$ 
of the continuum states. 
For these quantities we have derived, see Appendix~\ref{appendixA}, the following 
coupled system of NEGF equations of motion
\be
\left\{
\begin{array}{l}
\dot{\r}=-i\left[h_{\rm HF}[\r],\r\right]
-\callI[\r,f]-\callI^{\dag}[\r,f]
\\ \\
\dot{f}_{\m}=-\callJ_{\m}[\r,f]-\callJ^{\ast}_{\m}[\r,f]
\end{array}
\right..
\label{CHEERSeq}
\ee
The  generalized collision 
integral $\callI[\r,f]$ is defined as in Eq.~(\ref{collwoAuger}) but $\S[\r]\to 
\S[\r]+\S_{\rm Auger}[\r,f]$. The Auger self-energy is calculated 
from the second-order (in $v^{A}$) diagrams, in accordance with 
Refs.~\cite{PhysRevB.39.3489,PhysRevB.39.3503}, and reads
\bea
\S^{\lessgtr}_{{\rm Auger},ij}(t,\bar{t})=
\sum_{mn\, pq}\sum_{\m}G_{mn}^{\lessgtr}(t,\bar{t})
\nn\\
\times
\left[G_{\m}^{\lessgtr}(t,\bar{t})G^{\gtrless}_{pq}(\bar{t},t)
(v^{A}_{iqm\m}w^{A}_{\m 
npj}+v^{A}_{iq\m m}w^{A}_{n\m pj})
\right.
\nn\\
+\left.
G_{pq}^{\lessgtr}(t,\bar{t}) G_{\m}^{\gtrless}(\bar{t},t)
v^{A}_{i\m pm}w^{A}_{nq\m j}\right],
\label{2bse-Auger}
\eea
where we neglected 
the off-diagonal elements of the continuum Green's 
function, i.e., $G^{\lessgtr}_{\m\n}=\d_{\m\n}G^{\lessgtr}_{\m}$. As 
we shall demonstrate this approximation is remarkably accurate. 
Through the GKBA, 
$\S_{\rm Auger}$ is a time-nonlocal functional of $\r$ and $f_{\m}$.
Finally, the 
collision integral $\callJ_{\m}$ reads
\be
\callJ_{\m}(t)=
\int_{0}^{t} \!d\bar{t}\left[
K^{>}_{\m\m}(t,\bar{t})f^{<}_{\m}(\bar{t})
+K^{<}_{\m\m}(t,\bar{t})f^{>}_{\m}(\bar{t})\right],
\label{Jint}
\ee
where the kernel 
\bea
K^{\lessgtr}_{\m\n}(t,\bar{t})&=&i\sum_{mn\, pq\,sr}
v^{A}_{\m rpm}w^{A}_{nqs\n}
\nn\\
&\times&\!\!
G^{\lessgtr}_{mn}(t,\bar{t})G^{\lessgtr}_{pq}(t,\bar{t})G^{\gtrless}_{sr}(\bar{t},t)
e^{-i\e_{\n}(\bar{t}-t)}\quad
\eea
is a time-nonlocal functional  of $\r$ only. 
Equations~(\ref{CHEERSeq}), together with the definitions that follow 
it, constitute the first (methodological) result of this Letter.
The implementation of Eqs.~(\ref{CHEERSeq}) does not alter the quadratic scaling 
with $N_{t}$. The scaling with the number of basis functions 
changes from $N_{\rm bound}^{\mathfrak{p}}$ to 
$\max[N_{\rm bound}^{\mathfrak{p}},N_{\rm bound}^{\mathfrak{q}}N_{\rm cont}]$ 
where $N_{\rm cont}$ is the number of continuum states and 
$2\leq\mathfrak{q}\leq 4$. Therefore, the proposed equations can be 
used to simulate a large class of molecules of current interest.

{\em Assessment of NEGF approach}:
To demonstrate the reliability of the coupled NEGF Eqs.~(\ref{CHEERSeq}) we consider a 
1D atom with soft Coulomb interactions. 
On the grid points $x_{n}=na$ with $|n|<N_{\rm grid}/2$, the
single-particle Hamiltonian reads $h(x_{n},x_{m})=\d_{n,m}[2\k+V_{\rm 
n}(x_{n})]-\d_{|n-m|,1}\k$, where the nuclear potential $V_{\rm n}(x)=U_{\rm 
en}/\sqrt{x^{2}+a^{2}}$ for $|x|\leq R$ and $V_{\rm n}(x)=0$ 
otherwise. Electrons interact only in a box of length $2R$ 
centered around zero through $v(x,x')=ZU_{\rm 
ee}/\sqrt{(x-x')^{2}+a^{2}}$. The coupling to an 
external laser pulse is accounted for by adding $\d_{nm}x_{n}E(t)$ to 
$h(x_{n},x_{m})$. 

We take $N_{\rm grid}=400$
and (henceforth all quantities are expressed 
in atomic units) $a=0.5$,  $\k=2$, $Z=4$, $U_{\rm en}=2$, $U_{\rm 
ee}=U_{\rm en}/2$ and $R=10a$. 
With four electrons the HF spectrum has $N_{\rm sys}=5$  bound 
states (per spin) and $N_{\rm cont}=N_{\rm grid}-N_{\rm sys}$ continuum states. 
The occupied levels have energy  $\e_{c}=-4.33$ (core) and 
$\e_{v}=-1.65$ 
(valence).
The HF states are used to construct the 
Hamiltonian in Eqs.~(\ref{eqham}) and (\ref{laserham}). 
The results obtained by solving the coupled NEGF Eqs.~(\ref{CHEERSeq}) 
[where $\r$ is a $N_{\rm sys}\times N_{\rm sys}$ matrix and $f$ is a 
$N_{\rm cont}$--dimensional vector] are benchmarked against 
NEGF calculations on the {\em full} grid (NEGF$@$grid). 
NEGF$@$grid simulations are performed by solving 
the original equation~\cite{PUvLS.2015}
$\dot{\r}=-i\left[h_{\rm HF}[\r],\r\right]
-\callI[\r]-\callI^{\dag}[\r]$ where all quantities are 
$N_{\rm grid}\times N_{\rm grid}$ matrices in the $x_{n}$--basis 
and $\callI$ is given by Eq.~(\ref{collwoAuger}) with $\S=\S_{\rm 
c}$, see Appendix~\ref{appendixB} for details. 
By construction, NEGF$@$grid simulations include the off-diagonal 
elements $h_{i\m},\;h_{\m\m'},\;\bld_{\m\m'}$
and all Coulomb integrals with two or more indices in the continuum.
Notice that NEGF$@$grid scales cubically with $N_{\rm cont}$ and it 
is therefore not exportable to large systems.

\begin{figure}[tbp]
\includegraphics[width=0.49\textwidth]{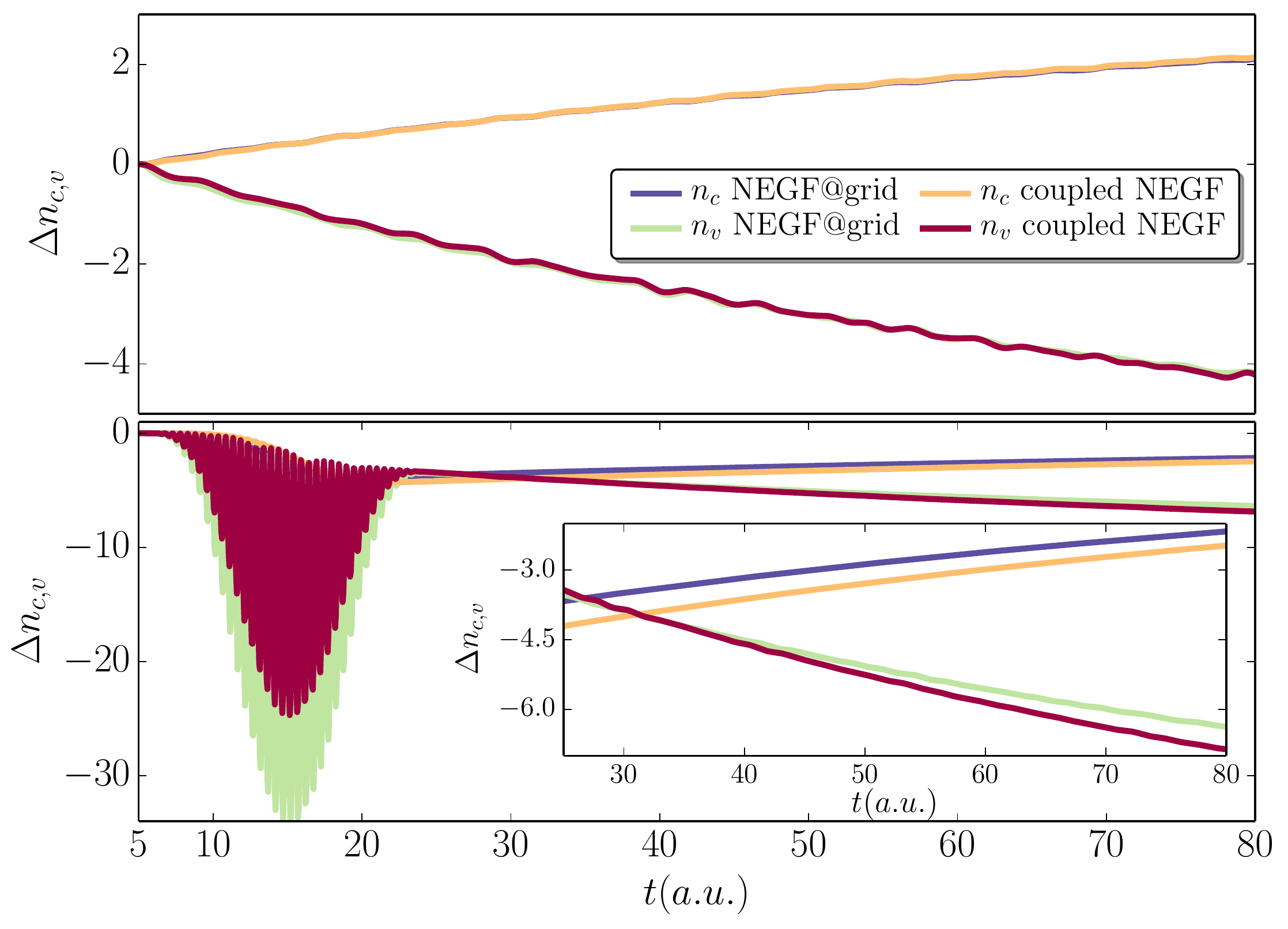}
\caption{Variation of the TD occupations (per spin) $n_{c}(t)$ [core, 
increasing blue (dark gray) and orange (light gray) curves] and $n_{v}(t)$ 
[valence, decreasing green (light gray) and red (dark gray) 
curves] calculated using NEGF$@$grid
and coupled NEGF Eqs.~(\ref{CHEERSeq}) for the sudden 
creation of a core hole (top) and the action of a laser pulse 
(bottom). The inset shows a magnification of  $n_{c}(t)$ and 
$n_{v}(t)$ after the end of the pulse. Vertical axes have been scaled 
up by a factor $10^{2}$.}
\label{ncnv}
\end{figure} 

In Fig.~\ref{ncnv} we show the time-dependent (TD) occupation (per 
spin) of the core, $n_{c}$, and valence, $n_{v}$, levels.
In the top panel we suddenly remove 4$\%$ of charge 
from the core, hence $\r_{cc}\to \r_{cc}-0.04$, and let the system 
evolve {\em without} external fields. In the bottom panel the 
equilibrium system is driven by the external pulse
\be
E(t)=E_{0}\sin^{2}\left(\frac{\p t}{T}\right)\sin(\W t)
\label{laser}
\ee
with central frequency 
$\W=6.2$, active from $t=0$ until $t=T=20$. 
The frequency is large enough for the energy of the photoelectron  
not to overlap with the energy of the Auger electron.
The intensity has been chosen to have the same amount of 
expelled charge as in the case of the sudden removal: $E_{0}=2.0$ for NEGF$@$grid 
and $E_{0}=1.5$ for the coupled NEGF Eqs.~(\ref{CHEERSeq}) 
-- the difference in the value of $E_{0}$ is due 
to the neglect of the dipole  elements 
$d_{\m\m'}$ in Eq.~(\ref{laserham}). 
The results 
perfectly agree in the top panel whereas only a minor discrepancy is observed 
in the bottom panel. 
In both type of simulations the Auger decay  slightly depends on how the core hole is 
created. In fact, the laser pulse is also responsible for expelling 
charge from the valence level, thereby hindering the refilling of the 
core. The core-hole lifetime  agrees well with the 
inverse linewidth function $\G(\e_{\rm Auger})=2\p\sum_{\m}|v_{c\m 
vv}|^{2}\d(\e_{\rm Auger}-\e_{\m})\simeq 10^{-2}$ in all cases. It is worth emphasizing that 
 no time-local approximation of $\S_{\rm Auger}$ would yield the 
behavior $n_{c}(t)=1-n_{h}e^{-\G t}$.  
We performed TD HF simulations both in the grid basis and by 
solving 
Eqs.~(\ref{CHEERSeq}) with $\callJ_{\m}=\S_{\rm c}=\S_{\rm 
Auger}=0$, and found that $n_{c}(t)$ remains essentially constant 
(not shown). This is consistent with similar findings obtained in TDDFT using 
adiabatic xc potentials~\cite{PhysRevB.86.045114}.

\begin{figure}[tbp]
\includegraphics[width=0.51\textwidth]{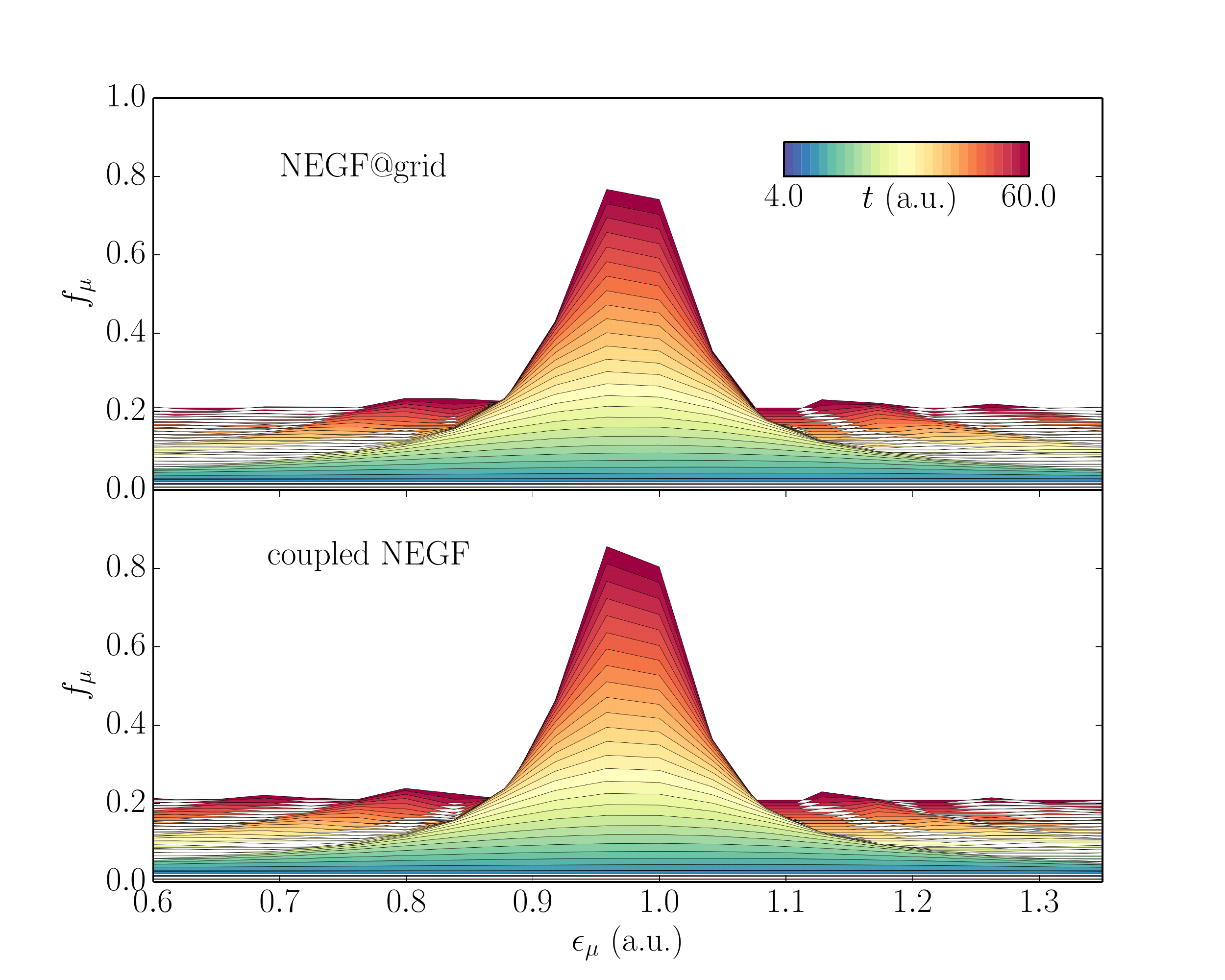}
\caption{Time-dependent occupations $f_{\m}(t)$ of the continuum states versus 
their energy $\e_{\m}$ after the sudden creation of a core hole.
The results are  
obtained from the solution of the NEGF$@$grid equation (top) 
and coupled NEGF Eqs.~(\ref{CHEERSeq}) (bottom). 
In both cases the maximum occurs at $\e_{\m}=\e_{\rm Auger}$.
Vertical axes have been scaled up by
a factor $10^{2}$. }
\label{fmu}
\end{figure} 

After the sudden creation of a core-hole 
the electronic density  populates the continuum states $\vf_{\m}$. 
In Fig.~\ref{fmu} we show the corresponding time-dependent 
occupations $f_{\m}(t)$ versus their energy $\e_{\m}$. 
Again simulations have been performed using NEGF$@$grid (top 
panel) and the 
coupled NEGF Eqs.~(\ref{CHEERSeq}) (bottom panel). 
As time passes the total expelled charge increases and 
$f_{\m}(t)$ gets peaked at the Auger energy 
$\e_{\rm Auger}=2\e_{v}-\e_{c}\simeq 1$. 
The final profile of the peak has a width $\G\equiv\G(\e_{\rm Auger})$, independently of how the 
core hole is created (suddenly or due to a laser pulse). On the contrary, the 
photoelectron peak attains a width  $\sim 2\p/T$
immediately after the end of the pulse.
We also observe that the exact 
energy of the Auger electron $\e^{\rm exact}_{\rm Auger}=2\e_{v}-\e_{c}-v_{vvvv}$ is not 
within reach the second-order approximation in Eq.~(\ref{2bse-Auger}):  the 
shift $v_{vvvv}$ (due to the valence--valence repulsion) would require a $T$-matrix 
treatment~\cite{cini1993two,PhysRevLett.39.504}. 
However, such shift has only a minor impact on the internal 
dynamics of 3D systems like, e.g., organic molecules, since 
the repulsion between two valence holes is typically less than 1 eV.

{\em Auger wavepacket reconstruction:}
We now use the coupled NEGF Eqs.~(\ref{CHEERSeq}) to study
the 1D atom on larger boxes (hence one-- and two--electron integrals 
are calculated from  HF states that spread over a large number of grid 
points). The output has been postprocessed to reconstruct the 
density of the Auger wavepacket according to $n_{\rm 
Auger}(x,t)=\sum_{\m\n}\vf_{\m}^{\ast}(x)f_{\m\n}(t)\vf_{\n}(x)$, 
where $f_{\m\n}(t)=\bra\hat{c}^{\dag}_{\n\s}(t)\hat{c}_{\m\s}(t)\ket$ 
is the off-diagonal density matrix in the continuum sector.
The latter is obtained by integrating the NEGF equation of motion 
(see SM for the derivation)
\be
\dot{f}_{\m\n}=-i(\e_{\m}-\e_{\n})f_{\m\n}-\callJ_{\m\n}[\r,f]
-\callJ^{\ast}_{\n\m}[\r,f],
\ee
where $\callJ_{\m\n}$ is given by the right hand side of Eq.~(\ref{Jint}) 
after the replacement $K^{\lessgtr}_{\m\m}(t,\bar{t})f_{\m}^{\gtrless}(\bar{t})\to
K^{\lessgtr}_{\m\n}(t,\bar{t})f_{\n}^{\gtrless}(\bar{t})$.

\begin{figure}[tbp]
\includegraphics[width=0.49\textwidth]{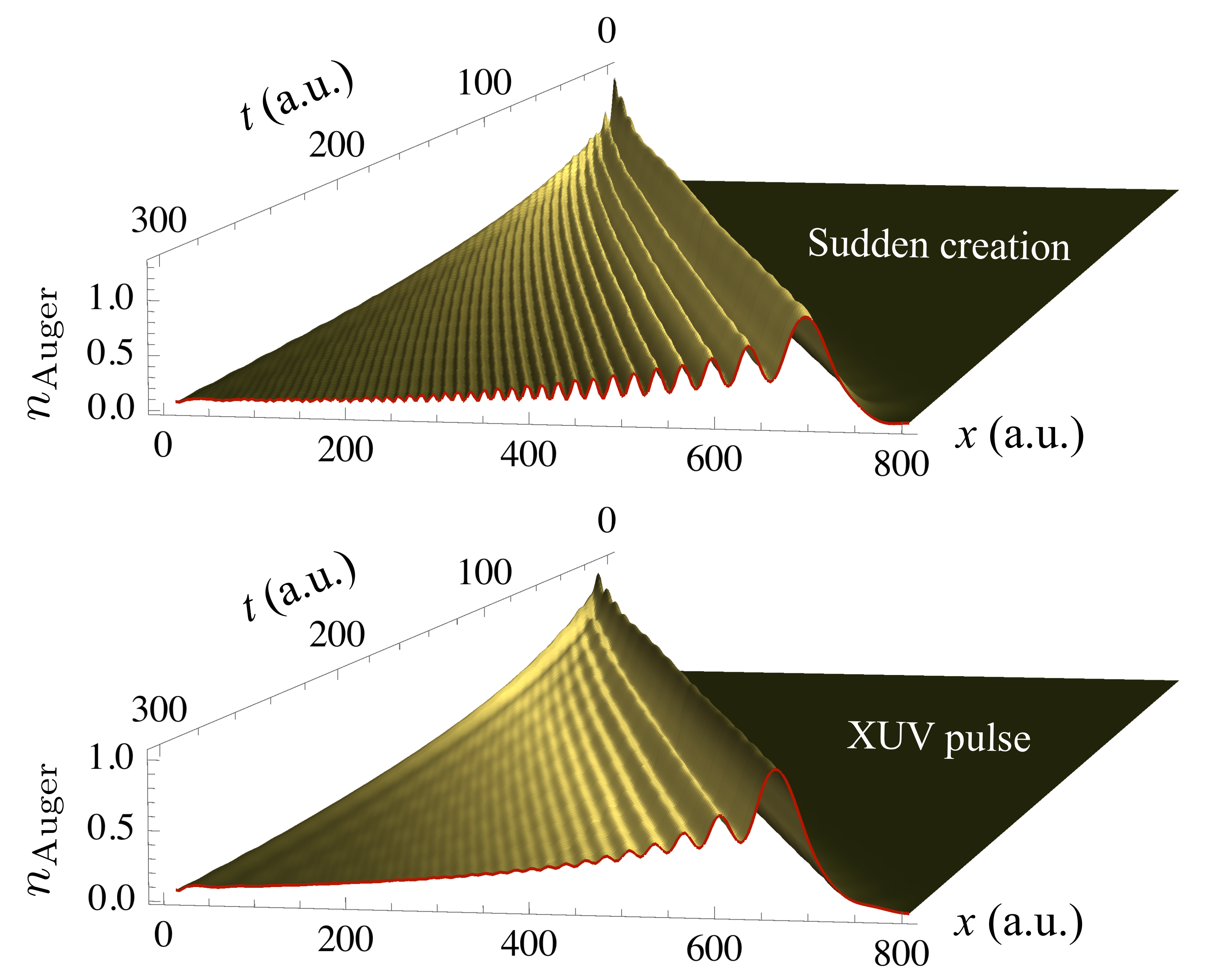}

\vspace{0.2cm}
\includegraphics[width=0.48\textwidth]{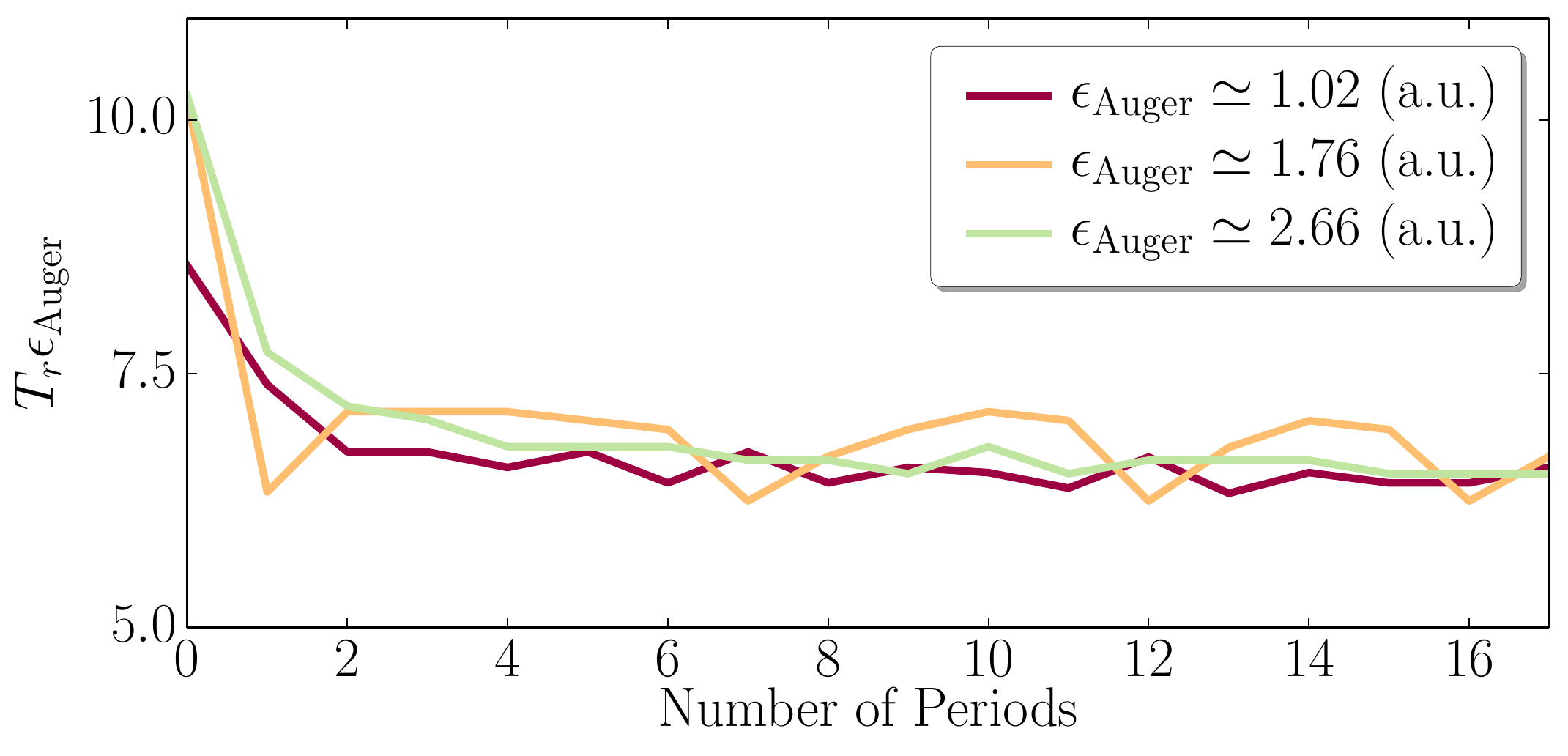}
\caption{Snapshots of the density of the Auger wavepacket after the 
sudden creation of a core hole (top) and the action of a laser pulse 
(middle). The bottom panel shows the period of the ripples at an 
interface versus the number of periods for three different values of range and strengths of 
the Coulomb force (see main text) yielding Auger energies $\e_{\rm 
Auger}=1.02$ (red -- dark gray), $1.76$ (yellow -- gray) and $2.66$ 
(green -- light gray).}
\label{wavepacket}
\end{figure}

In Fig.~\ref{wavepacket} we display the Auger wavepacket 
for $N_{\rm grid}=1600$ grid-points. In the top 
panel the core hole is suddenly created
whereas in the middle panel 
the atom is driven by the ionizing laser of Eq.~(\ref{laser}). The 
first observation is that the wavefront depends on the perturbation 
(sudden creation or laser), being steeper the shorter it takes to 
create the hole.
The wavepacket moves rightward at the expected speed $v= \de\e/\de p \simeq 
2\sqrt{\k\e_{\rm Auger}}=2.2$ and its length is approximately 
$v/\G$ far away from the nucleus. Interestingly, the  
tail of the wavepacket exhibits spatial ripples that tend to accumulate 
nearby the origin. The amplitude of the ripples depends on the perturbation 
(sudden creation or laser) whereas their spacing is an {\em intrinsic} feature. In the bottom 
panel of Fig.~\ref{wavepacket} we show the period $T_{r}$ of the 
ripples, i.e., the elapsing time between two consecutive 
maxima of $n_{\rm Auger}(x_{0},t)$, at 
the interface $x_{0}=30 a$,  versus the number of periods.
We present results for three different values of range and strengths of 
the Coulomb force 
$(R,U_{\rm en},U_{\rm ee})=
(10a,2,1),\;(100a,2.6,2.08),\;(10a,2.7,2.025)$
yielding Auger energies $\e_{\rm 
Auger}=1.02,\;1.76,\;2.66$ respectively. 
In all cases we find that 
$T_{r}$  attains a finite limit given by
\be
T_{r}=2\p/\e_{\rm Auger}.
\label{Augerperiod}
\ee
The occurrence of ripples and the intrinsic period $T_{r}$ is not an 
artifact of the self-energy approximation. These features as well as the overall 
shape of the Auger wavepacket are indeed 
confirmed by CI calculations.
Starting at 
time $t=0$ with the 
photoexcited state $|\F_{\rm x}\ket=\hat{c}^{\dag}_{c\ua}\hat{c}^{\dag}_{v\da}
\hat{c}^{\dag}_{v\ua}|0\ket$ and evolving with the Hamiltonian in 
Eq.~(\ref{eqham})
one finds $n_{\rm Auger}(x,t)=|\vf_{\rm Auger}(x,t)|^{2}$ 
with $\vf_{\rm Auger}(x,t)=\sum_{\m}a_{\m}(t)\vf_{\m}(x)$ and 
\be
a_{\m}(t)\simeq -v_{c\m vv}e^{-iE_{\m}t}\;
\frac{e^{i(\e_{\m}-\e_{\rm Auger}+i\G/2)t}-1}{\e_{\m}-\e_{\rm 
Auger}+i\G/2}.
\label{ciripple}
\ee
The CI Auger wavepacket is in excellent agreement with NEGF, see 
Appendix~\ref{appendixC}. 
In the Appendix~\ref{appendixC} we further show that the  
ripples occur even in two or three dimensions and, therefore, they 
are a fingerprint of the Auger electron.

To summarize, we have included Auger 
decays in a  
first-principles NEGF approach 
to simulate the UCM dynamics of molecules driven by 
attosecond pulses. The computational effort is comparable to that of 
previous NEGF implementations~\cite{PUvLS.2015,C60paper2018,PSMP.2018}, 
thereby allowing for studying systems with 
tens of active electrons up to tens of femtoseconds. Benchmarks 
in 1D atoms demonstrate that both qualitative and quantitative aspects 
are well captured. We also predict a highly asymmetric profile of the 
Auger  wavepacket with a spatial extension of the 
order $v/\G$
and superimposed ripples with temporal period $T_{r}=2\p/\e_{\rm Auger}$.

Although the fundamental equations have been derived for finite systems, 
the proposed NEGF approach can be extended to deal with 
periodic systems too. In this context the 
equation of motion for the single particle density matrix 
opens the possibility to 
develop current-density functional theories that include dissipation 
and thermalization.

{\em Akcknowledgements}
G.S. and E.P. acknowledge EC funding through the RISE Co-ExAN (Grant No. GA644076).
E.P. also acknowledges funding from the European Union project 
MaX Materials design at the eXascale H2020-EINFRA-2015-1, Grant Agreement No.
676598 and Nanoscience Foundries and
Fine Analysis-Europe H2020-INFRAIA-2014-2015, Grant Agreement No. 654360.
F.C and A.R. 
acknowledge financial support from the European Research Council 
(ERC-2015-AdG-694097), Grupos Consolidados (IT578-13) and European 
Union Horizon 2020  program under Grant Agreement  676580 (NOMAD).

\appendix

\section{Derivation of NEGF equations in HF basis}
\label{appendixA}

The starting point is the equation of motion for the Green's function 
$\callG(z,z')$ with times $z,z'$ on the Keldysh contour. 
For the 
Hamiltonian in Eqs.~(1) and~(2)
 it is convenient to write $\callG$ and the correlation self-energy 
$\calgS$ in a block form
\be
\callG(z,z')=\left(\begin{array}{cc}
G(z,z') & \D(z,z') \\ \bar{\D}(z,z') & C(z,z')
\end{array}\right),
\ee
\be
\calgS(z,z')=\left(\begin{array}{cc}
\S_{G}(z,z') & \S_{\D}(z,z') \\ \bar{\S}_{\D}(z,z') & \S_{C}(z,z')
\end{array}\right),
\ee
where $G$ is a matrix with indices in the bound sector, $C$ is a 
matrix with indices in the continuum sector and $\D$, $\bar{\D}$ are 
the off-diagonal blocks. The blocks of the self-energy have the same 
structure. For the self-energy we make the following approximation
\\

(i) All self-energy diagrams containing $\D$ or $\bar{\D}$ 
propagators are set to zero (see below for the justification).
\\

From the approximation (i) it follows that $\S_{\D}=\bar{\S}_{\D}=0$ 
and that the 
Hartree-Fock (HF) potential has indices only in the bound sector 
since the Coulomb integrals in $\hat{H}^{\rm eq}$  have at most one 
index in the continuum. The explicit form of the HF potential is
\be
V_{{\rm HF},ij}(z)=-i\sum_{mn}G_{nm}(z,z^{+})w_{imnj},
\ee
where $w_{imnj}\equiv 2v_{imnj}-v_{imjn}$.

The equations of motion for the different blocks of $\callG$ 
then read (in matrix form)
\bea
\left[i\frac{d}{dz}-h_{{\rm HF}}(z)\right]\!G(z,z')
-\left(\blE(z)\cdot\bld\right) \bar{\D}(z,z')
\nn\\
=\d(z,z')+\int d\bar{z}\;\S_{G}(z,\bar{z})G(\bar{z},z')
\label{eom1}
\eea
\bea
\left[i\frac{d}{dz}-\callE\right]\!\bar{\D}(z,z')
-\left(\blE(z)\cdot\bld\right) G(z,z')
\nn\\
=\d(z,z')+\int d\bar{z}\;\S_{C}(z,\bar{z})\bar{\D}(\bar{z},z')
\label{eom2}
\eea
\bea
\left[i\frac{d}{dz}-\callE\right]\!C(z,z')
=\d(z,z')+\int d\bar{z}\;\S_{C}(z,\bar{z})C(\bar{z},z')
\label{eom3}
\eea
where in Eq.~(\ref{eom1}) we have defined the nonequilibrium 
single-particle HF Hamiltonian 
\be
h_{{\rm HF}}=h+V_{\rm HF}+\blE\cdot\bld,
\ee
and in the last two equations we have defined the matrix 
$\callE_{\m\n}=\d_{\m\n}\e_{\m}$. The blocks of the dipole matrix are 
unambiguously determined by the contractions and we do therefore use the 
same symbol for all four blocks. Notice that no coupling with the 
electric field appears in Eq.~(\ref{eom3}) since we set 
$\bld_{\m\m'}=0$ in Eq.~(2).

Next we observe that if the energy-window of the photoelectron does 
not overlap with that of the 
Auger electron then we can make the 
approximation:
\\

(ii) $\S_{C}(z,\bar{z})\bar{\D}(\bar{z},z')\simeq 0$.
\\

With the approximation (ii) we easily integrate Eq.~(\ref{eom2}) and 
obtain 
\be
\bar{\D}_{\m j}(z,z')=\sum_{n}\int d\bar{z}\;C^{0}_{\m}(z,\bar{z})
\left(\blE(\bar{z})\cdot\bld_{\m n}\right) G_{nj}(\bar{z},z'),
\label{solbard}
\ee
where $C^{0}$ is the solution of Eq.~(\ref{eom3}) with $\S_{C}=0$. 
Since $\callE$ is diagonal so is $C^{0}$.

Inserting Eq.~(\ref{solbard}) into Eq.~(\ref{eom1}) we get
\bea
\left[i\frac{d}{dz}-h_{{\rm HF}}(z)\right]\!G(z,z')
=\d(z,z')\quad\quad\quad
\nn\\
+\int d\bar{z}\left[\S_{G}(z,\bar{z})+\S_{\rm 
ion}(z,\bar{z})\right]G(\bar{z},z'),
\label{eom1v2}
\eea
where we have defined the ionization self-energy
\be
\S_{{\rm ion},ij}(z,\bar{z})\equiv \sum_{\m}
\left(\blE(z)\cdot\bld_{i\m}\right)C^{0}_{\m}(z,\bar{z})
\left(\blE(\bar{z})\cdot\bld_{\m j}\right).
\ee
The diagrammatic representation of the ionization self-energy is 
displayed in the bottom diagram of Fig.~1(b)  where, to avoid a 
proliferation of different symbols, we used $G^{0}_{\m}$ instead of 
$C^{0}_{\m}$  (in the main text we also used $G_{\m\n}$ instead of
$C_{\m\n}$). Notice that $\S_{{\rm 
ion}}$ vanishes for times at which the external pulse is zero.

We now have to specify the approximation for the correlation 
self-energy. For weakly interacting closed systems (no 
continuum states) the 
self-consistent second-Born approximation (2B) has been shown to be 
accurate in several nonequilibrium 
situations~\cite{PhysRevA.82.033427,BalzerHermanns2012,Sakkinen-2012,HermannsPRB2014,CTPPBonitz2016,HopjanPRL2016,ReichmanEPL2016,Joost2017,UKSSKvLG.2011,C60paper2018}. 
The very same approximation describes Auger scatterings provided that 
we also consider 
interaction lines with one index in the 
continuum~\cite{PhysRevB.39.3489,PhysRevB.39.3503}.  We therefore 
approximate $\S_{G}$ and $\S_{C}$ as the sum of the 2B diagrams. 
It is easy to show that for a $\callG$  initially block diagonal (no 
electrons in the continuum in the ground state) the off-diagonal 
blocks remain zero for all times in the 2B 
approximation. This justifies the 
approximation (i). 

\begin{figure}[tbp]
\includegraphics[width=0.43\textwidth]{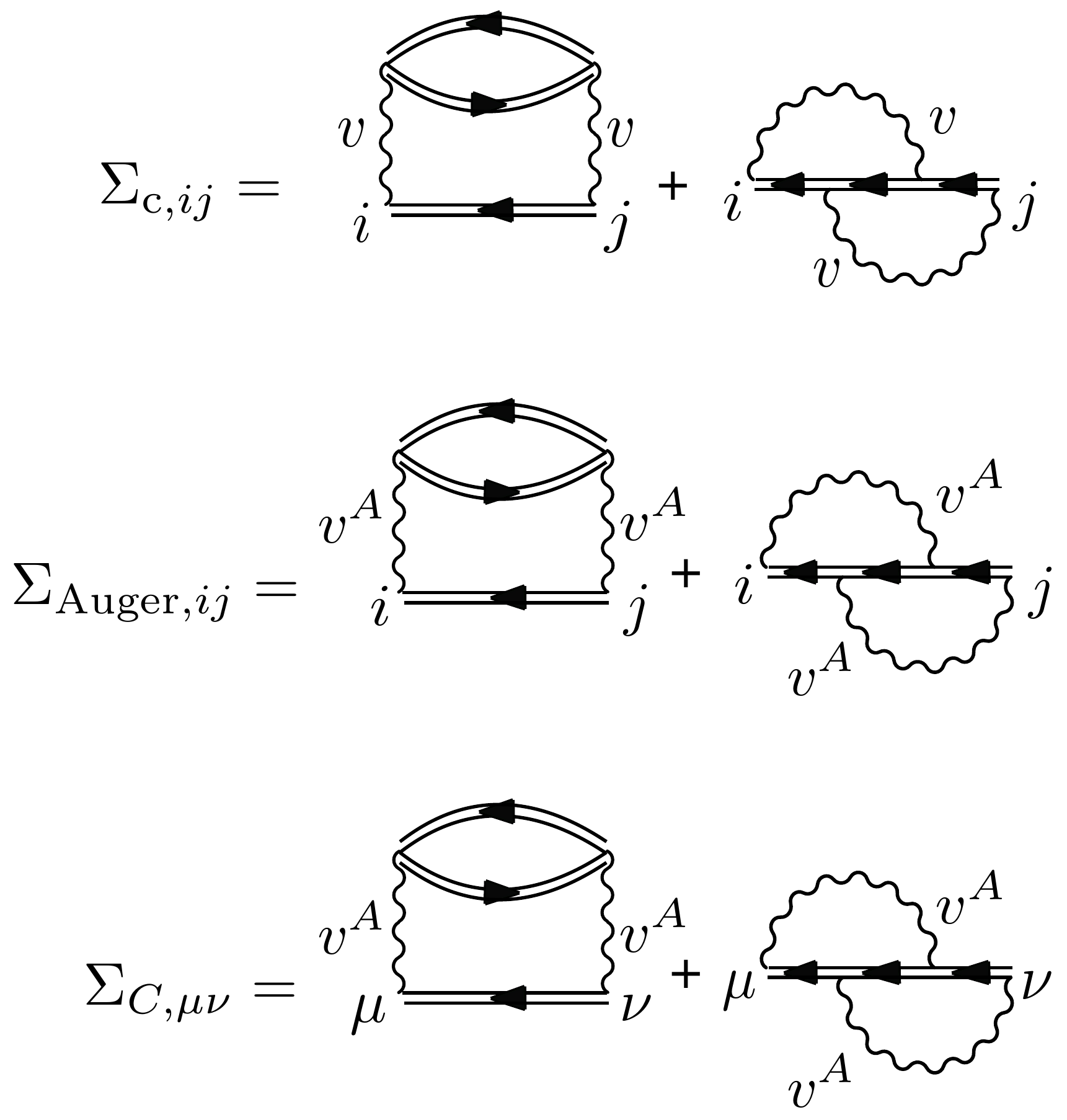}
\caption{Self-energy diagrams with indices in the bound sector 
for intramolecular scattering (top) and Auger scattering (middle).  
Self-energy diagrams for Auger electrons (bottom).}
\label{sigmas}
\end{figure} 

The 2B diagrams for $\S_{G}$ can be split into diagrams with
interaction lines having all indices in the bound sector ($v$) 
and diagrams with interaction lines having one index in the continuum 
sector ($v^{A}$):
\be
\S_{G}=\S_{\rm c}+\S_{\rm Auger}.
\ee
Using the Feynman rules, see top and middle panel of 
Fig.~\ref{sigmas}, one finds
\bea
\S_{{\rm c},ij}(z,z')&=&
\sum_{mn,pq,sr}v_{irpm}w_{nqsj}
\nn\\
&\times&
G_{mn}(z,z')
G_{pq}(z,z')G_{sr}(z',z),
\label{2bse}
\eea
and 
\bea
\S_{{\rm Auger},ij}(z,z')=
\sum_{mn\, pq}\sum_{\m}G_{mn}(z,z')
\nn\\
\times
\left[C_{\m\n}(z,z')G_{pq}(z',z)
(v^{A}_{iqm\m}w^{A}_{\n 
npj}+v^{A}_{iq\m m}w^{A}_{n\n pj})
\right.
\nn\\
+\left.
G_{pq}(z,z') C_{\m\n}(z',z)
v^{A}_{i\n pm}w^{A}_{nq\m j}\right].
\label{2bse-Auger}
\eea
The correlation self-energy $\S_{\rm c}$ is also given in the top 
diagram of Fig.~1(b).

The 2B diagrams for $\S_{C}$ do instead contain only $v^{A}$ interaction 
lines since both indices of $\S_{C}$  are in the continuum sector. From the 
bottom diagram of Fig.~\ref{sigmas} one finds
\bea
\S_{C,\m\n}(z,z')&=&
\sum_{mn,pq,sr}v^{A}_{\m rpm}w^{A}_{nqs\n}
\nn\\
&\times&
G_{mn}(z,z')
G_{pq}(z,z')G_{sr}(z',z).
\label{2bse-cont}
\eea

For a short and weak laser pulse the off-diagonal matrix elements of 
$C$ are small. We therefore make the approximation
\\

(iii) $C_{\m\n}\simeq \d_{\m\n}C_{\m}$ in $\S_{{\rm Auger}}$
\\

Implementing (iii) in Eq.~(\ref{2bse-Auger}) and extracting the 
lesser/greater component we get precisely the self-energy in 
Eq.~(5).

To summarize, with the approximations (i-iii) the equations of 
motion become
\bea
\left[i\frac{d}{dz}-h_{{\rm HF}}(z)\right]\!G(z,z')
=\d(z,z')+\int d\bar{z}\;\S(z,\bar{z})G(\bar{z},z')
\nn\\
\label{eomG}
\eea
\bea
\left[i\frac{d}{dz}-\callE\right]\!C(z,z')
=\d(z,z')+\int d\bar{z}\;\S_{C}(z,\bar{z})C(\bar{z},z')
\nn\\
\label{eomC}
\eea
where in Eq.~(\ref{eomG}) we have defined 
\be
\S\equiv\S_{\rm c}+\S_{\rm ion}+\S_{\rm Auger}.
\ee

Taking the adjoint of Eqs.~(\ref{eomG},\ref{eomC}), summing the 
resulting equations to 
Eqs.~(\ref{eomG},\ref{eomC}) and evaluating the result in 
$z=z^{+}=t$ we get the equation of motion for the density matrices 
$\r_{ij}(t)=-iG_{ij}(z,z^{+})$ and $f_{\m\n}(t)=-iC_{\m\n}(z,z^{+})$:
\bea 
\dot{\r}=-i\left[h_{\rm HF},\r\right]
-\callI-\callI^{\dag},
\label{rhodot}
\eea
\be
\dot{f}_{\m\n}=-i(\e_{\m}-\e_{\n})f_{\m\n}-\callJ_{\m\n}
-\callJ^{\ast}_{\n\m},
\label{effedot}
\ee
where
\be
\callI(t)=
\int_{0}^{t}\! d\bar{t}\!\left[\S^{>}(t,\bar{t})G^{<}(\bar{t},t)-
\S^{<}(t,\bar{t})G^{>}(\bar{t},t)\right],
\label{collwoAuger}
\ee
\be
\callJ(t)=
\int_{0}^{t}\! d\bar{t}\!\left[\S^{>}_{C}(t,\bar{t})C^{<}(\bar{t},t)-
\S_{C}^{<}(t,\bar{t})C^{>}(\bar{t},t)\right].
\label{collwoAugerJ}
\ee

Equations~(\ref{rhodot},\ref{effedot}) do not close on $\r$ 
and $f$ since the right hand side depends on $G$ and $C$ calculated 
at different times. To close the equations we make the 
Generalized Kadanoff-Baym Ansatz~\cite{PhysRevB.34.6933} (GKBA). According to the GKBA we can replace 
all $G^{\lessgtr}$ and $C^{\lessgtr}$ appearing in $\callI$ and $\callJ$ with
\be
G^{\lessgtr}(t,\bar{t})=\mp
\left[G^{\rm R}(t,t')\r^{\lessgtr}(t')-\r^{\lessgtr}(t)G^{\rm A}(t,t')\right],
\ee
\be
C^{\lessgtr}(t,\bar{t})=\mp
\left[C^{\rm R}(t,t')f^{\lessgtr}(t')-f^{\lessgtr}(t)C^{\rm A}(t,t')\right],
\ee
where $\r^{<}=\r$, $\r^{>}=1-\r$ and similarly $f^{<}=f$, $f^{>}=1-f$. 
For the retarded/advanced Green's function we consider the HF 
approximation according to which
\be
G^{\rm R}(t,t')=[G^{\rm A}(t',t)]^{\dag}=-i\th(t-t')\callT
\left[e^{-i\int_{t'}^{t}d\bar{t}\,h_{\rm HF}(\bar{t})}\right],
\ee
\be
C^{\rm R}_{\m\n}(t,t')=[C^{\rm A}_{\n\m}(t',t)]^{\ast}=
-i\d_{\m\n}\th(t-t')e^{-i\e_{\m}(t-t')}.
\ee
Since $h_{\rm HF}$ is a functional of $\r$ we see that 
Eqs.~(\ref{rhodot},\ref{effedot}) become nonlinear integro-differential 
equations for $\r_{ij}(t)$ and $f_{\m\n}(t)$. Notice also that in the 
equation for $\r$ the dependence on $f$ is only through the diagonal 
elements $f_{\m}\equiv 
f_{\m\m}$ appearing in $\S_{\rm Auger}$, due to the approximation 
(iii). If we set $\m=\n$ in Eq.~(\ref{effedot}) then for the right 
hand side to depend only on $f_{\m}$ we have to make the approximation
\\

(iv) $f_{\m\n}=\d_{\m\n}f_{\m}$ in $\callJ$
\\

\noindent
which is consistent with the approximation (iii).

It is easy to show that in this way the equation for $\r$ becomes 
the first of Eqs.~(4) and that the equation for $f_{\m\n}$ 
becomes Eq.~(9), which for $\m=\n$ reduces to the second of Eqs.~(4).

\section{NEGF$@$grid versus coupled NEGF calculations}
\label{appendixB}

To assess the accuracy of the approximations made at the level of the 
Hamiltonian with Eqs.~(1,2) and at the level of NEGF with (i-iv),
we considered a 1D atom on a grid. In the grid basis 
the total Hamiltonian in second quantization reads
\bea
\hat{H}(t)&=&\sum_{\substack{mn\\ \s}}
\q^{\dag}_{\s}(x_{m})h(x_{m},x_{n})\q_{\s}(x_{n})
\nn\\
&+&\frac{1}{2}\sum_{\substack{mn\\ \s\s'}}\q^{\dag}_{\s}(x_{m})\q^{\dag}_{\s'}(x_{n})v(x_{m},x_{n})
\q_{\s'}(x_{n})\q_{\s}(x_{m})
\nn\\
&+&E(t)\sum_{\substack{m\\ 
\s}}x_{m}\q^{\dag}_{\s}(x_{m})\q_{\s}(x_{m}).
\label{gridham}
\eea
where the one-particle Hamiltonian $h(x,x')$ and the interaction 
$v(x,x')$ are defined in the main text.
The equation of motion for the density matrix in grid basis
$\r(x_{m},x_{n},t)=G(x_{m},z;x_{n},z^{+})$ in the 2B 
approximation is
\bea
\dot{\r}(x_{m},x_{n},t)=-i\sum_{p}\left[
h_{\rm HF}(x_{m},x_{p},t)\r(x_{p},x_{n},t)
\right.
\nn\\\left.
-\r(x_{m},x_{p},t)h_{\rm HF}(x_{p},x_{n},t)\right]
\nn\\
-\callI_{g}(x_{m},x_{n},t)-\callI_{g}^{\ast}(x_{n},x_{m},t).
\label{rdotgrid}
\eea
In Eq.~(\ref{rdotgrid}) we have the HF Hamiltonian in  
grid basis
\be
h_{\rm HF}(x_{m},x_{p},t)=h(x_{m},x_{p})
+V_{\rm HF}(x_{m},x_{p},t)+\d_{mp} E(t)x_{m},
\label{hHFgrid}
\ee
with HF potential
\bea
V_{\rm HF}(x_{m},x_{p},t)
&=&2\d_{nm}\sum_{q}v(x_{m},x_{q})\r(x_{q},x_{q},t)
\nn\\
&-&v(x_{m},x_{p})\r(x_{m},x_{p},t),
\eea
and the collision integral in  grid basis
\bea
\callI_{g}(x_{m},x_{n},t)=\sum_{p}
\int_{0}^{t}\! 
d\bar{t}\!\left[\S_{g}^{>}(x_{m},t;x_{p},\bar{t})G^{<}(x_{p},\bar{t};x_{n},t)\right.
\nn\\
\left.
\S^{<}_{g}(x_{m},t;x_{p},\bar{t})G^{>}(x_{p},\bar{t};x_{n},t)\right],
\nn\\
\eea
with the 2B self-energy 
\bea
\S_{g}^{\lessgtr}(x_{m},t;x_{p},\bar{t})=
\sum_{rs}v(x_{m},x_{r})v(x_{p},x_{s})
\nn\\
\times\left[2G^{\lessgtr}(x_{m},t;x_{p},\bar{t})
G^{\lessgtr}(x_{r},t;x_{s},\bar{t})
G^{\gtrless}(x_{s},\bar{t};x_{r},t)\right.
\nn\\
-\left.
G^{\lessgtr}(x_{m},t;x_{s},\bar{t})G^{\gtrless}(x_{s},\bar{t};x_{r},t)
G^{\lessgtr}(x_{r},t;x_{p},\bar{t})\right].
\eea

The NEGF$@$grid results have been obtained by solving 
Eq.~(\ref{rdotgrid}) with lesser/greater Green's function evaluated 
at the GKBA level. Except that for the 2B approximation to $\S_{g}$, 
no other approximation has been made. For a 
system with $N_{\rm grid}$ points this require to propagate and store 
matrices $N_{\rm grid}\times N_{\rm grid}$.

In order to apply the coupled NEGF scheme based on Eqs.~(4) we first  
solve the self-consistent HF problem and 
extract the equilibrium bound eigenfunctions $\vf_{i}(x_{n})$ 
and continuum eigenfunctions $\vf_{\m}(x_{n})$ of
energy $\e_{i}$ and $\e_{\m}$ respectively. The HF eigenfunctions are 
then used to calculate the matrix elements in the bound sector 
of the one-particle Hamiltonian
\be
h_{ij}=\sum_{mn}\vf^{\ast}_{i}(x_{m})h(x_{m},x_{n})\vf_{j}(x_{m}),
\ee
the dipole operator
\be
d_{ij}=\sum_{m}\vf^{\ast}_{i}(x_{m})x_{m}\vf_{j}(x_{m}),
\ee
and the Coulomb repulsion
\be
v_{ijpq}=\sum_{mn}\vf^{\ast}_{i}(x_{m})\vf^{\ast}_{j}(x_{n})
v(x_{m},x_{n})\vf_{p}(x_{n})\vf_{q}(x_{m}).
\ee
The continuum HF eigenfunctions are used to calculate the 
bound-continuum matrix elements of the dipole operator
\be
d_{i\m}=\sum_{m}\vf^{\ast}_{i}(x_{m})x_{m}\vf_{\m}(x_{m}),
\ee
and the  Coulomb repulsion responsible for Auger scatterings 
\be
v^{A}_{ijp\m}=\sum_{mn}\vf^{\ast}_{i}(x_{m})\vf^{\ast}_{j}(x_{n})
v(x_{m},x_{n})\vf_{p}(x_{n})\vf_{\m}(x_{m}).
\ee

With this information we approximate the original Hamiltonian in 
Eq.~(\ref{gridham}) in accordance with Eqs.~(1,2), i.e.,
\bea
\hat{H}(t)=\sum_{\substack{ij\\ \s}}h_{ij}\hat{c}^{\dag}_{i\s}\hat{c}_{j\s}
+\frac{1}{2}\sum_{\substack{ ijpq\\ \s\s'}}
v_{ijpq}\hat{c}^{\dag}_{i\s}\hat{c}^{\dag}_{j\s'}\hat{c}_{p\s'}\hat{c}_{q\s}
\nn\\
+\sum_{\m\s}\e_{\m}\hat{c}^{\dag}_{\m\s}\hat{c}_{\m\s}+\sum_{\substack{ ijp\m\\ \s\s'}}
v^{A}_{ijp\m}
\left(\hat{c}^{\dag}_{i\s}\hat{c}^{\dag}_{j\s'}\hat{c}_{p\s'}\hat{c}_{\m\s}
+{\rm h.c.}\right)
\nn\\
+E(t)
\sum_{\substack{ij\\ \s}}d_{ij}\hat{c}^{\dag}_{i\s}\hat{c}_{j\s}
+E(t)\sum_{\substack{i\m\\ 
\s}}\left(d_{i\m}\hat{c}^{\dag}_{i\s}\hat{c}_{\m\s}+{\rm 
h.c.}\right),
\label{MBhamHFbasis}
\eea
where $\hat{c}_{i\s}$ ($\hat{c}_{\m\s}$) are annihilation operators for an electron in 
the HF orbital $\vf_{i}$ ($\vf_{\m}$) with spin $\s$. Of course, had we included 
in Eq.~(\ref{MBhamHFbasis}) the off-diagonal one-electron terms containing 
$h_{i\m}$, $h_{\m\m'}$ and $d_{\m\m'}$ and the interaction terms 
containing $v_{ij\m\m'}$, $v_{i\n\m\m'}$   and $v_{\n'\n\m\m'}$ we 
would have got the same Hamiltonian as in Eq.~(\ref{gridham}) but in the 
HF basis.

With the approximate Hamiltonian in Eq.~(\ref{MBhamHFbasis}) 
we solve the coupled NEGF equations 
(4) which, we emphasize again, have been derived by making   
the additional approximations (i-iv) of the previous section. The agreement 
between the full-grid simulations and the simulations based on 
Eqs.~(4)  indicate that the latter are enough to capture
qualitatively and quantitatively the physics of the Auger decay.

We observe that in the grid simulations the self-energy $\S_{g}$ 
contains all possible scatterings, including those contained in the 
self-energies $\S_{\rm c}$ and $\S_{\rm Auger}$ of the  coupled NEGF 
scheme. Furthermore, in the grid simulations no ionization 
self-energy appears since the photoionization is  
accounted for by explicitly including all grid points (even those 
far away from the nucleus). In other words, {\em all} elements 
$\r(x_{m},x_{n},t)$ are coupled and propagated in time.

\section{CI versus coupled NEGF calculations}
\label{appendixC}

To further check the quality of the NEGF  Eqs.~(4) we have also 
solved the time-dependent problem using a Configuration Interaction 
(CI) expansion. 

The neutral 1D atom described in the main text of the paper has 
four electrons, two in the core and two in the valence levels. 
We are interested in suddenly removing a core electron of, say, spin 
down, and in studying how 
the system evolves with the Hamiltonian in 
Eq.~(\ref{MBhamHFbasis}). For the CI expansion we use the following 
three-body states
\bea
|\F_{x}\ket&=&\hat{c}^{\dag}_{c\ua}\hat{c}^{\dag}_{v\da}\hat{c}^{\dag}_{v\ua}|0\ket,
\\
|\F_{g}\ket&=&\hat{c}^{\dag}_{c\ua}\hat{c}^{\dag}_{c\da}\hat{c}^{\dag}_{v\ua}|0\ket,
\\
|\F_{\m}\ket&=&\hat{c}^{\dag}_{c\ua}\hat{c}^{\dag}_{c\da}\hat{c}^{\dag}_{\m\ua}|0\ket,
\eea
describing the initially photoionized state ($\F_{x}$), the cationic 
ground state ($\F_{g}$) and the Auger states ($\F_{\m}$). We expand 
the state of the system at time $t$ according to
\be
|\Q(t)\ket=a_{x}(t)|\F_{x}\ket+a_{g}(t)|\F_{g}\ket+
\sum_{\m}a_{\m}(t)|\F_{\m}\ket,
\ee
and impose the initial condition $a_{x}(0)=1$ and 
$a_{g}(0)=a_{\m}(0)=0$. Using the fact that in the HF basis 
$h_{\rm HF}$ is diagonal, it is easy to show that the cationic ground 
state decouples and the dynamics is governed by the equations
below
\bea
i\dot{a}_{x}&=&E_{x}a_{x}+\sum_{\m}v_{c\m vv}a_{\m},
\label{dotax}
\\
i\dot{a}_{\m}&=&v_{c\m vv}a_{x}+E_{\m}a_{\m}.
\label{dotamu}
\eea
The three-body energies are
\bea
E_{x}&=&2\e_{v}+\e_{c}-v_{cccc}-4v_{cvvc}+2v_{cvcv}-v_{vvvv},\quad
\label{Ex}
\\
E_{\m}&=&\e_{\m}+2\e_{c}-v_{cccc}-4v_{cvvc}+2v_{cvcv},
\eea
where the HF energies of the core and valence levels are given by
\bea
\e_{c}&=&h_{cc}+v_{cccc}+2v_{cvvc}-v_{cvcv},
\\
\e_{v}&=&h_{vv}+v_{vvvv}+2v_{vccv}-v_{vcvc}.
\eea

The energy $\e_{\rm Auger}=\e_{\m_{A}}$ of the Auger electron is determined by the condition 
$E_{\m_{A}}=E_{x}$ which yields
\be
\e_{\rm Auger}=2\e_{v}-\e_{c}-v_{vvvv}
\ee
as it should. The red-shift $v_{vvvv}$ is due to the repulsion of the 
two holes in the final state. In order to capture this red-shift using 
Many-Body Perturbation Theory (MBPT) one should go beyond the 2B 
approximation for the self-energy and consider the $T$-matrix 
approximation in the particle-particle 
sector~\cite{cini1993two,PhysRevLett.39.504}. We observe, however, 
that for weakly correlated molecules, like organic molecules and 
biomolecules, the magnitude of 
the valence-valence repulsion is typically less than 1 eV; hence, 
neglecting this repulsion does not 
substantially affect the dynamics during the first ten of femtoseconds 
or so. 
\begin{figure}[tbp]
\includegraphics[width=0.49\textwidth]{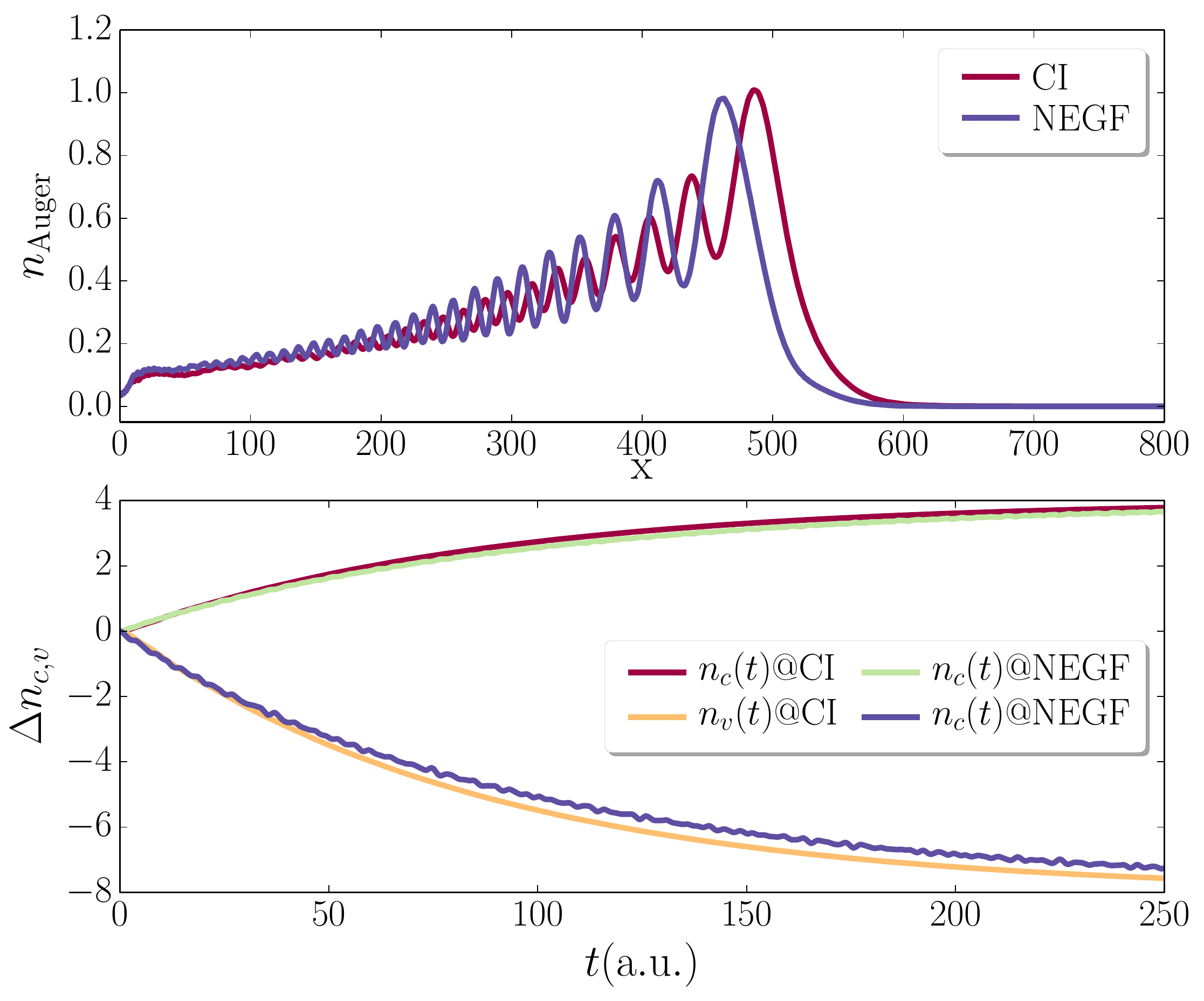}
\caption{Auger wavepacket (top) and variation of the occupations of 
the core and valence levels (bottom) in CI and in coupled NEGF. Same 
parameters as in top panel of Fig.~4.}
\label{NEGF-CI}
\end{figure}

For the 1D atom the valence-valence repulsion is mainly responsible for 
reducing the speed of the Auger electron. The form of the Auger 
wavepacket as well as the time-dependent behavior of the refilling of 
the core-hole are not altered if we set $v_{vvvv}=0$ in Eq.~(\ref{Ex}). 
For a fair comparison with the coupled 
NEGF Eqs.~(4) we therefore solve Eqs.~(\ref{dotax},\ref{dotamu}) 
using $E_{x}^{2B}=E_{x}+v_{vvvv}$ in place of $E_{x}$. In 
Fig.~\ref{NEGF-CI} we compare the Auger wavepacket (top panel) and 
the occupation of the core and valence levels (bottom panels) 
calculated using CI and the coupled NEGF equations (4). Also in 
this case the 
agreement is rather satisfactory.

\begin{figure}[tbp]
\includegraphics[width=0.43\textwidth]{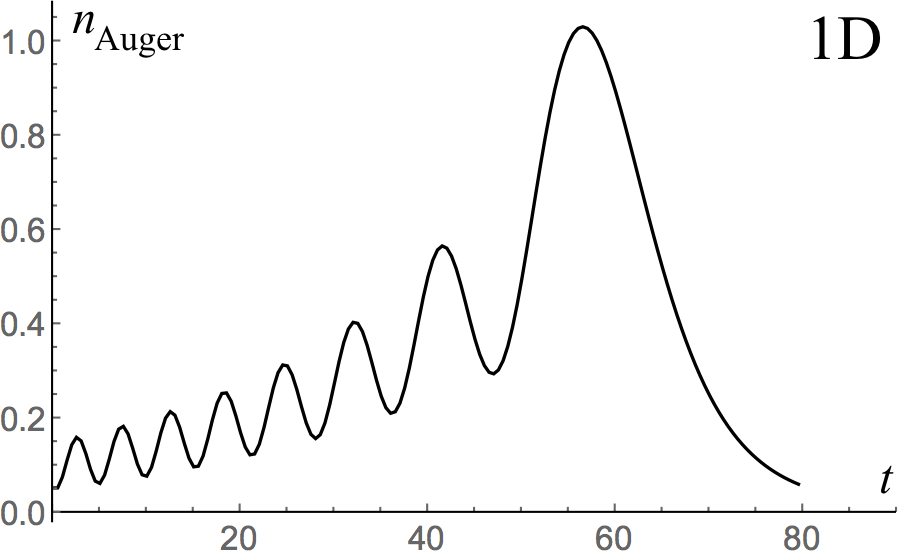}
\includegraphics[width=0.43\textwidth]{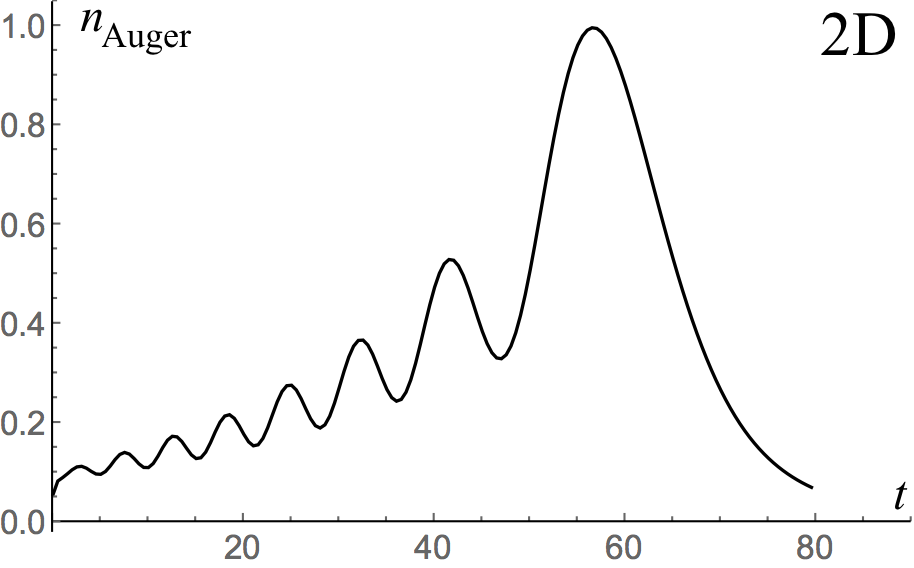}
\includegraphics[width=0.43\textwidth]{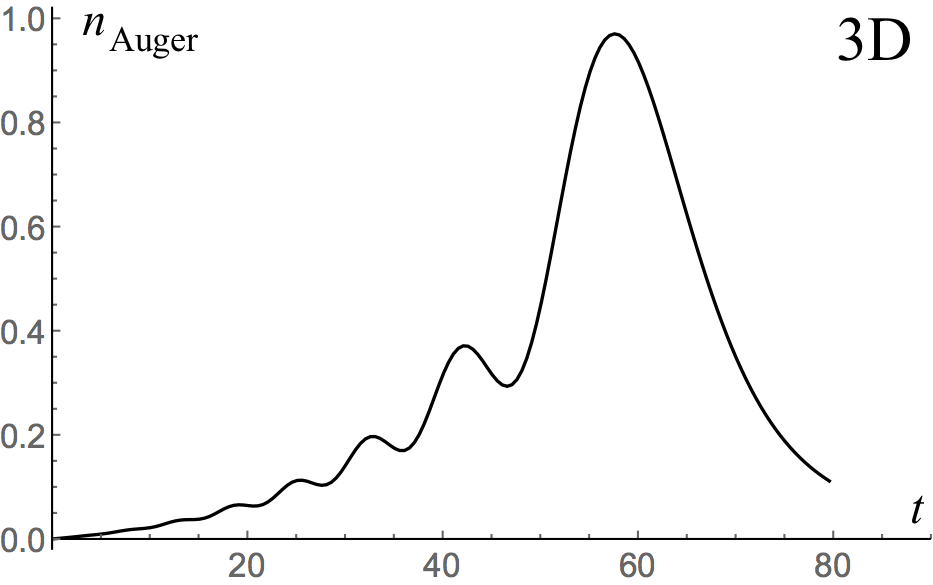}
\caption{Auger wavepacket (in arbitrary units) for $\G=0.05$ and $\e_{\rm Auger}=1$ after 
a time $t=50$ from the sudden removal of a core electron. Top: 
$n_{\rm Auger}(r,t)$ in 1D. Middle: $rn_{\rm Auger}(r,t)$ in 2D. 
Bottom: $r^{2}n_{\rm Auger}(r,t)$ in 3D.}
\label{ripplesfig}
\end{figure} 

The analytic calculation can be carried on further if we assume that 
the broadening
\be
\G(\w)=2\p\sum_{\m}|v_{c\m vv}|^{2}\d(\w-\e_{\m})
\ee
is a weakly dependent function of $\w$ for $\w\simeq \e_{\rm Auger}$. 
In this case it is straightforward to show that the 
amplitudes $a_{\m}$ are given by
\be
a_{\m}(t)=-v_{c\m vv}e^{-iE_{\m}t}\;
\frac{e^{i(\e_{\m}-\e_{\rm Auger}+i\G/2)t}-1}{\e_{\m}-\e_{\rm Auger}+i\G/2}
\label{ciripple}
\ee
which coincides with Eq.~(11). The occurrence of ripples on the 
tail of the Auger wavepacket stems from 
the structure of the $a_{\m}$'s. In fact, the ripples are independent of the dimension of the system
and of the details of the continuum states in the vicinity of the 
nucleus. 
As an example, let $\m=\blp$ be the momentum in D dimension and 
let us use planewaves $\vf_{\m}(\blr)=\vf_{\blp}(\blr)=e^{i\blp\cdot\blr}$ 
for the continuum states. We further consider a free dispersion 
$\e_{\m}=\e_{\blp}=p^{2}/2$ and, for simplicity, an Auger interaction 
$v_{c\m vv}=v_{c\blp vv}$ independent of $\blp$ so that 
$a_{\m}=a_{p}$ depends only on the modulus $p=|\blp|$ of the 
momentum, see Eq.~(\ref{ciripple}). Then, the 
Auger wavepacket is spherically symmetric and its density is given by
\be
n_{\rm Auger}(r,t)=\left|
\int \frac{d^{D}p}{(2\p)^{D}}a_{p}(t)e^{i\blp\cdot\blr}
\right|^{2}.
\ee
In Fig.~\ref{ripplesfig} we show $n_{\rm Auger}(r,t)$ for
$\G=0.05$ and an Auger energy $\e_{\rm Auger}=1$ after a time 
$t=50$ from the sudden removal of the core electron. The figure shows 
$n_{\rm Auger}(r,t)$ in 1D (top), $rn_{\rm Auger}(r,t)$ in 2D 
(middle) and $r^{2}n_{\rm Auger}(r,t)$ in 3D (bottom). 
In all cases we appreciate the occurrence of ripples although they 
tend to get smeared out as the dimension increases.


\begin{thebibliography}{42}
\expandafter\ifx\csname natexlab\endcsname\relax\def\natexlab#1{#1}\fi
\expandafter\ifx\csname bibnamefont\endcsname\relax
  \def\bibnamefont#1{#1}\fi
\expandafter\ifx\csname bibfnamefont\endcsname\relax
  \def\bibfnamefont#1{#1}\fi
\expandafter\ifx\csname citenamefont\endcsname\relax
  \def\citenamefont#1{#1}\fi
\expandafter\ifx\csname url\endcsname\relax
  \def\url#1{\texttt{#1}}\fi
\expandafter\ifx\csname urlprefix\endcsname\relax\def\urlprefix{URL }\fi
\providecommand{\bibinfo}[2]{#2}
\providecommand{\eprint}[2][]{\url{#2}}

\bibitem[{\citenamefont{Calegari et~al.}(2014)\citenamefont{Calegari, Ayuso,
  Trabattoni, Belshaw, De~Camillis, Anumula, Frassetto, Poletto, Palacios,
  Decleva et~al.}}]{Calegari336}
\bibinfo{author}{\bibfnamefont{F.}~\bibnamefont{Calegari}},
  \bibinfo{author}{\bibfnamefont{D.}~\bibnamefont{Ayuso}},
  \bibinfo{author}{\bibfnamefont{A.}~\bibnamefont{Trabattoni}},
  \bibinfo{author}{\bibfnamefont{L.}~\bibnamefont{Belshaw}},
  \bibinfo{author}{\bibfnamefont{S.}~\bibnamefont{De~Camillis}},
  \bibinfo{author}{\bibfnamefont{S.}~\bibnamefont{Anumula}},
  \bibinfo{author}{\bibfnamefont{F.}~\bibnamefont{Frassetto}},
  \bibinfo{author}{\bibfnamefont{L.}~\bibnamefont{Poletto}},
  \bibinfo{author}{\bibfnamefont{A.}~\bibnamefont{Palacios}},
  \bibinfo{author}{\bibfnamefont{P.}~\bibnamefont{Decleva}},
  \bibnamefont{et~al.}, \bibinfo{journal}{Science}
  \textbf{\bibinfo{volume}{346}}, \bibinfo{pages}{336} (\bibinfo{year}{2014}),
  ISSN \bibinfo{issn}{0036-8075},
  \urlprefix\url{http://science.sciencemag.org/content/346/6207/336}.

\bibitem[{\citenamefont{Uiberacker et~al.}(2007)\citenamefont{Uiberacker,
  Uphues, Schultze, Verhoef, Yakovlev, Kling, Rauschenberger, Kabachnik,
  Schr{\"o}der, Lezius et~al.}}]{uiberacker2007attosecond}
\bibinfo{author}{\bibfnamefont{M.}~\bibnamefont{Uiberacker}},
  \bibinfo{author}{\bibfnamefont{T.}~\bibnamefont{Uphues}},
  \bibinfo{author}{\bibfnamefont{M.}~\bibnamefont{Schultze}},
  \bibinfo{author}{\bibfnamefont{A.~J.} \bibnamefont{Verhoef}},
  \bibinfo{author}{\bibfnamefont{V.}~\bibnamefont{Yakovlev}},
  \bibinfo{author}{\bibfnamefont{M.~F.} \bibnamefont{Kling}},
  \bibinfo{author}{\bibfnamefont{J.}~\bibnamefont{Rauschenberger}},
  \bibinfo{author}{\bibfnamefont{N.~M.} \bibnamefont{Kabachnik}},
  \bibinfo{author}{\bibfnamefont{H.}~\bibnamefont{Schr{\"o}der}},
  \bibinfo{author}{\bibfnamefont{M.}~\bibnamefont{Lezius}},
  \bibnamefont{et~al.}, \bibinfo{journal}{Nature}
  \textbf{\bibinfo{volume}{446}}, \bibinfo{pages}{627} (\bibinfo{year}{2007}).

\bibitem[{\citenamefont{Kuleff and Cederbaum}(2014)}]{KuleffCederbaum}
\bibinfo{author}{\bibfnamefont{A.~I.} \bibnamefont{Kuleff}} \bibnamefont{and}
  \bibinfo{author}{\bibfnamefont{L.~S.} \bibnamefont{Cederbaum}},
  \bibinfo{journal}{Journal of Physics B: Atomic, Molecular and Optical
  Physics} \textbf{\bibinfo{volume}{47}}, \bibinfo{pages}{124002}
  (\bibinfo{year}{2014}),
  \urlprefix\url{http://stacks.iop.org/0953-4075/47/i=12/a=124002}.

\bibitem[{\citenamefont{Kuleff et~al.}(2016)\citenamefont{Kuleff, Kryzhevoi,
  Pernpointner, and Cederbaum}}]{PhysRevLett.117.093002}
\bibinfo{author}{\bibfnamefont{A.~I.} \bibnamefont{Kuleff}},
  \bibinfo{author}{\bibfnamefont{N.~V.} \bibnamefont{Kryzhevoi}},
  \bibinfo{author}{\bibfnamefont{M.}~\bibnamefont{Pernpointner}},
  \bibnamefont{and} \bibinfo{author}{\bibfnamefont{L.~S.}
  \bibnamefont{Cederbaum}}, \bibinfo{journal}{Phys. Rev. Lett.}
  \textbf{\bibinfo{volume}{117}}, \bibinfo{pages}{093002}
  (\bibinfo{year}{2016}),
  \urlprefix\url{https://link.aps.org/doi/10.1103/PhysRevLett.117.093002}.

\bibitem[{\citenamefont{Breidbach and Cederbaum}(2003)}]{BreidbachCederbaum}
\bibinfo{author}{\bibfnamefont{J.}~\bibnamefont{Breidbach}} \bibnamefont{and}
  \bibinfo{author}{\bibfnamefont{L.~S.} \bibnamefont{Cederbaum}},
  \bibinfo{journal}{The Journal of Chemical Physics}
  \textbf{\bibinfo{volume}{118}}, \bibinfo{pages}{3983} (\bibinfo{year}{2003}),
  \eprint{http://dx.doi.org/10.1063/1.1540618},
  \urlprefix\url{http://dx.doi.org/10.1063/1.1540618}.

\bibitem[{\citenamefont{Golubev and Kuleff}(2015)}]{PhysRevA.91.051401}
\bibinfo{author}{\bibfnamefont{N.~V.} \bibnamefont{Golubev}} \bibnamefont{and}
  \bibinfo{author}{\bibfnamefont{A.~I.} \bibnamefont{Kuleff}},
  \bibinfo{journal}{Phys. Rev. A} \textbf{\bibinfo{volume}{91}},
  \bibinfo{pages}{051401} (\bibinfo{year}{2015}),
  \urlprefix\url{https://link.aps.org/doi/10.1103/PhysRevA.91.051401}.

\bibitem[{\citenamefont{Nagaya et~al.}(2010)\citenamefont{Nagaya, Iwayama,
  Sugishima, Ohmasa, and Yao}}]{Nagaya.2010}
\bibinfo{author}{\bibfnamefont{K.}~\bibnamefont{Nagaya}},
  \bibinfo{author}{\bibfnamefont{H.}~\bibnamefont{Iwayama}},
  \bibinfo{author}{\bibfnamefont{A.}~\bibnamefont{Sugishima}},
  \bibinfo{author}{\bibfnamefont{Y.}~\bibnamefont{Ohmasa}}, \bibnamefont{and}
  \bibinfo{author}{\bibfnamefont{M.}~\bibnamefont{Yao}},
  \bibinfo{journal}{Applied Physics Letters} \textbf{\bibinfo{volume}{96}},
  \bibinfo{pages}{233101} (\bibinfo{year}{2010}),
  \eprint{https://doi.org/10.1063/1.3442483},
  \urlprefix\url{https://doi.org/10.1063/1.3442483}.

\bibitem[{\citenamefont{Kutzner et~al.}(2002)\citenamefont{Kutzner, Maycock,
  Thorarinson, Pannwitz, and Robertson}}]{PhysRevA.66.042715}
\bibinfo{author}{\bibfnamefont{M.}~\bibnamefont{Kutzner}},
  \bibinfo{author}{\bibfnamefont{V.}~\bibnamefont{Maycock}},
  \bibinfo{author}{\bibfnamefont{J.}~\bibnamefont{Thorarinson}},
  \bibinfo{author}{\bibfnamefont{E.}~\bibnamefont{Pannwitz}}, \bibnamefont{and}
  \bibinfo{author}{\bibfnamefont{J.~A.} \bibnamefont{Robertson}},
  \bibinfo{journal}{Phys. Rev. A} \textbf{\bibinfo{volume}{66}},
  \bibinfo{pages}{042715} (\bibinfo{year}{2002}),
  \urlprefix\url{https://link.aps.org/doi/10.1103/PhysRevA.66.042715}.

\bibitem[{\citenamefont{Uphues et~al.}(2008)\citenamefont{Uphues, Schultze,
  Kling, Uiberacker, Hendel, Heinzmann, Kabachnik, and Drescher}}]{Uphues2008}
\bibinfo{author}{\bibfnamefont{T.}~\bibnamefont{Uphues}},
  \bibinfo{author}{\bibfnamefont{M.}~\bibnamefont{Schultze}},
  \bibinfo{author}{\bibfnamefont{M.~F.} \bibnamefont{Kling}},
  \bibinfo{author}{\bibfnamefont{M.}~\bibnamefont{Uiberacker}},
  \bibinfo{author}{\bibfnamefont{S.}~\bibnamefont{Hendel}},
  \bibinfo{author}{\bibfnamefont{U.}~\bibnamefont{Heinzmann}},
  \bibinfo{author}{\bibfnamefont{N.~M.} \bibnamefont{Kabachnik}},
  \bibnamefont{and} \bibinfo{author}{\bibfnamefont{M.}~\bibnamefont{Drescher}},
  \bibinfo{journal}{New Journal of Physics} \textbf{\bibinfo{volume}{10}},
  \bibinfo{pages}{025009} (\bibinfo{year}{2008}),
  \urlprefix\url{http://stacks.iop.org/1367-2630/10/i=2/a=025009}.

\bibitem[{\citenamefont{Drescher et~al.}(2002)\citenamefont{Drescher,
  Hentschel, Kienberger, Uiberacker, Yakovlev, Scrinzi, Westerwalbesloh,
  Kleineberg, Heinzmann, and Krausz}}]{drescher2002time}
\bibinfo{author}{\bibfnamefont{M.}~\bibnamefont{Drescher}},
  \bibinfo{author}{\bibfnamefont{M.}~\bibnamefont{Hentschel}},
  \bibinfo{author}{\bibfnamefont{R.}~\bibnamefont{Kienberger}},
  \bibinfo{author}{\bibfnamefont{M.}~\bibnamefont{Uiberacker}},
  \bibinfo{author}{\bibfnamefont{V.}~\bibnamefont{Yakovlev}},
  \bibinfo{author}{\bibfnamefont{A.}~\bibnamefont{Scrinzi}},
  \bibinfo{author}{\bibfnamefont{T.}~\bibnamefont{Westerwalbesloh}},
  \bibinfo{author}{\bibfnamefont{U.}~\bibnamefont{Kleineberg}},
  \bibinfo{author}{\bibfnamefont{U.}~\bibnamefont{Heinzmann}},
  \bibnamefont{and} \bibinfo{author}{\bibfnamefont{F.}~\bibnamefont{Krausz}},
  \bibinfo{journal}{Nature} \textbf{\bibinfo{volume}{419}},
  \bibinfo{pages}{803} (\bibinfo{year}{2002}), \bibinfo{note}{article},
  \urlprefix\url{http://dx.doi.org/10.1038/nature01143}.

\bibitem[{\citenamefont{Zherebtsov et~al.}(2011)\citenamefont{Zherebtsov,
  Wirth, Uphues, Znakovskaya, Herrwerth, Gagnon, Korbman, Yakovlev, Vrakking,
  Drescher et~al.}}]{zherebtsov2011attosecond}
\bibinfo{author}{\bibfnamefont{S.}~\bibnamefont{Zherebtsov}},
  \bibinfo{author}{\bibfnamefont{A.}~\bibnamefont{Wirth}},
  \bibinfo{author}{\bibfnamefont{T.}~\bibnamefont{Uphues}},
  \bibinfo{author}{\bibfnamefont{I.}~\bibnamefont{Znakovskaya}},
  \bibinfo{author}{\bibfnamefont{O.}~\bibnamefont{Herrwerth}},
  \bibinfo{author}{\bibfnamefont{J.}~\bibnamefont{Gagnon}},
  \bibinfo{author}{\bibfnamefont{M.}~\bibnamefont{Korbman}},
  \bibinfo{author}{\bibfnamefont{V.~S.} \bibnamefont{Yakovlev}},
  \bibinfo{author}{\bibfnamefont{M.}~\bibnamefont{Vrakking}},
  \bibinfo{author}{\bibfnamefont{M.}~\bibnamefont{Drescher}},
  \bibnamefont{et~al.}, \bibinfo{journal}{Journal of Physics B: Atomic,
  Molecular and Optical Physics} \textbf{\bibinfo{volume}{44}},
  \bibinfo{pages}{105601} (\bibinfo{year}{2011}).

\bibitem[{\citenamefont{Schins et~al.}(1994)\citenamefont{Schins, Breger,
  Agostini, Constantinescu, Muller, Grillon, Antonetti, and
  Mysyrowicz}}]{Schins-PhysRevLett.73.2180}
\bibinfo{author}{\bibfnamefont{J.~M.} \bibnamefont{Schins}},
  \bibinfo{author}{\bibfnamefont{P.}~\bibnamefont{Breger}},
  \bibinfo{author}{\bibfnamefont{P.}~\bibnamefont{Agostini}},
  \bibinfo{author}{\bibfnamefont{R.~C.} \bibnamefont{Constantinescu}},
  \bibinfo{author}{\bibfnamefont{H.~G.} \bibnamefont{Muller}},
  \bibinfo{author}{\bibfnamefont{G.}~\bibnamefont{Grillon}},
  \bibinfo{author}{\bibfnamefont{A.}~\bibnamefont{Antonetti}},
  \bibnamefont{and}
  \bibinfo{author}{\bibfnamefont{A.}~\bibnamefont{Mysyrowicz}},
  \bibinfo{journal}{Phys. Rev. Lett.} \textbf{\bibinfo{volume}{73}},
  \bibinfo{pages}{2180} (\bibinfo{year}{1994}),
  \urlprefix\url{https://link.aps.org/doi/10.1103/PhysRevLett.73.2180}.

\bibitem[{\citenamefont{Smirnova et~al.}(2003)\citenamefont{Smirnova, Yakovlev,
  and Scrinzi}}]{PhysRevLett.91.253001}
\bibinfo{author}{\bibfnamefont{O.}~\bibnamefont{Smirnova}},
  \bibinfo{author}{\bibfnamefont{V.~S.} \bibnamefont{Yakovlev}},
  \bibnamefont{and} \bibinfo{author}{\bibfnamefont{A.}~\bibnamefont{Scrinzi}},
  \bibinfo{journal}{Phys. Rev. Lett.} \textbf{\bibinfo{volume}{91}},
  \bibinfo{pages}{253001} (\bibinfo{year}{2003}),
  \urlprefix\url{https://link.aps.org/doi/10.1103/PhysRevLett.91.253001}.

\bibitem[{\citenamefont{Kazansky et~al.}(2009)\citenamefont{Kazansky, Sazhina,
  and Kabachnik}}]{Kazansky.2009}
\bibinfo{author}{\bibfnamefont{A.~K.} \bibnamefont{Kazansky}},
  \bibinfo{author}{\bibfnamefont{I.~P.} \bibnamefont{Sazhina}},
  \bibnamefont{and} \bibinfo{author}{\bibfnamefont{N.~M.}
  \bibnamefont{Kabachnik}}, \bibinfo{journal}{Journal of Physics B: Atomic,
  Molecular and Optical Physics} \textbf{\bibinfo{volume}{42}},
  \bibinfo{pages}{245601} (\bibinfo{year}{2009}),
  \urlprefix\url{http://stacks.iop.org/0953-4075/42/i=24/a=245601}.

\bibitem[{\citenamefont{Kazansky et~al.}(2011)\citenamefont{Kazansky, Sazhina,
  and Kabachnik}}]{Kazansky.2011}
\bibinfo{author}{\bibfnamefont{A.~K.} \bibnamefont{Kazansky}},
  \bibinfo{author}{\bibfnamefont{I.~P.} \bibnamefont{Sazhina}},
  \bibnamefont{and} \bibinfo{author}{\bibfnamefont{N.~M.}
  \bibnamefont{Kabachnik}}, \bibinfo{journal}{Journal of Physics B: Atomic,
  Molecular and Optical Physics} \textbf{\bibinfo{volume}{44}},
  \bibinfo{pages}{215601} (\bibinfo{year}{2011}),
  \urlprefix\url{http://stacks.iop.org/0953-4075/44/i=21/a=215601}.

\bibitem[{\citenamefont{Runge and Gross}(1984)}]{RungeGross:84}
\bibinfo{author}{\bibfnamefont{E.}~\bibnamefont{Runge}} \bibnamefont{and}
  \bibinfo{author}{\bibfnamefont{E.~K.~U.} \bibnamefont{Gross}},
  \bibinfo{journal}{Phys. Rev. Lett.} \textbf{\bibinfo{volume}{52}},
  \bibinfo{pages}{997} (\bibinfo{year}{1984}),
  \urlprefix\url{https://link.aps.org/doi/10.1103/PhysRevLett.52.997}.

\bibitem[{\citenamefont{Ullrich}(2012)}]{Ullrich:12}
\bibinfo{author}{\bibfnamefont{C.}~\bibnamefont{Ullrich}},
  \emph{\bibinfo{title}{Time-Dependent Density-Functional Theory}}
  (\bibinfo{publisher}{Oxford University Press}, \bibinfo{address}{Oxford},
  \bibinfo{year}{2012}).

\bibitem[{\citenamefont{Maitra}(2016)}]{Maitra.2016}
\bibinfo{author}{\bibfnamefont{N.~T.} \bibnamefont{Maitra}},
  \bibinfo{journal}{The Journal of Chemical Physics}
  \textbf{\bibinfo{volume}{144}}, \bibinfo{pages}{220901}
  (\bibinfo{year}{2016}), \eprint{https://doi.org/10.1063/1.4953039},
  \urlprefix\url{https://doi.org/10.1063/1.4953039}.

\bibitem[{\citenamefont{Cucinotta et~al.}(2012)\citenamefont{Cucinotta, Hughes,
  and Ballone}}]{PhysRevB.86.045114}
\bibinfo{author}{\bibfnamefont{C.~S.} \bibnamefont{Cucinotta}},
  \bibinfo{author}{\bibfnamefont{D.}~\bibnamefont{Hughes}}, \bibnamefont{and}
  \bibinfo{author}{\bibfnamefont{P.}~\bibnamefont{Ballone}},
  \bibinfo{journal}{Phys. Rev. B} \textbf{\bibinfo{volume}{86}},
  \bibinfo{pages}{045114} (\bibinfo{year}{2012}),
  \urlprefix\url{https://link.aps.org/doi/10.1103/PhysRevB.86.045114}.

\bibitem[{aug()}]{auger-tddft}
\bibinfo{note}{The inadequacy of the adiabatic approximation is easily
  understable. The Auger electron adds to the main quasi-particle peak a
  secondary peak in the spectral function of the parent cation. An adiabatic
  approximation can, at most, renormalize the main quasi-particle peak.}

\bibitem[{\citenamefont{Kadanoff and Baym}(1962)}]{kadanoff1962quantum}
\bibinfo{author}{\bibfnamefont{L.~P.} \bibnamefont{Kadanoff}} \bibnamefont{and}
  \bibinfo{author}{\bibfnamefont{G.~A.} \bibnamefont{Baym}},
  \emph{\bibinfo{title}{Quantum statistical mechanics: Green's function methods
  in equilibrium and nonequilibirum problems}} (\bibinfo{publisher}{Benjamin},
  \bibinfo{year}{1962}).

\bibitem[{\citenamefont{Stefanucci and van Leeuwen}(2013)}]{svl-book}
\bibinfo{author}{\bibfnamefont{G.}~\bibnamefont{Stefanucci}} \bibnamefont{and}
  \bibinfo{author}{\bibfnamefont{R.}~\bibnamefont{van Leeuwen}},
  \emph{\bibinfo{title}{Nonequilibrium Many-Body Theory of Quantum Systems: A
  Modern Introduction}} (\bibinfo{publisher}{Cambridge University Press},
  \bibinfo{address}{Cambridge}, \bibinfo{year}{2013}).

\bibitem[{\citenamefont{My\"oh\"anen et~al.}(2009)\citenamefont{My\"oh\"anen,
  Stan, Stefanucci, and van Leeuwen}}]{mssvl.2009}
\bibinfo{author}{\bibfnamefont{P.}~\bibnamefont{My\"oh\"anen}},
  \bibinfo{author}{\bibfnamefont{A.}~\bibnamefont{Stan}},
  \bibinfo{author}{\bibfnamefont{G.}~\bibnamefont{Stefanucci}},
  \bibnamefont{and} \bibinfo{author}{\bibfnamefont{R.}~\bibnamefont{van
  Leeuwen}}, \bibinfo{journal}{Phys. Rev. B} \textbf{\bibinfo{volume}{80}},
  \bibinfo{pages}{115107} (\bibinfo{year}{2009}),
  \urlprefix\url{https://link.aps.org/doi/10.1103/PhysRevB.80.115107}.

\bibitem[{\citenamefont{My\"oh\"anen et~al.}(2008)\citenamefont{My\"oh\"anen,
  Stan, Stefanucci, and van Leeuwen}}]{mssvl.2008}
\bibinfo{author}{\bibfnamefont{P.}~\bibnamefont{My\"oh\"anen}},
  \bibinfo{author}{\bibfnamefont{A.}~\bibnamefont{Stan}},
  \bibinfo{author}{\bibfnamefont{G.}~\bibnamefont{Stefanucci}},
  \bibnamefont{and} \bibinfo{author}{\bibfnamefont{R.}~\bibnamefont{van
  Leeuwen}}, \bibinfo{journal}{EPL (Europhysics Letters)}
  \textbf{\bibinfo{volume}{84}}, \bibinfo{pages}{67001} (\bibinfo{year}{2008}),
  \urlprefix\url{http://stacks.iop.org/0295-5075/84/i=6/a=67001}.

\bibitem[{\citenamefont{Latini et~al.}(2014)\citenamefont{Latini, Perfetto,
  Uimonen, van Leeuwen, and Stefanucci}}]{LPUvLS.2014}
\bibinfo{author}{\bibfnamefont{S.}~\bibnamefont{Latini}},
  \bibinfo{author}{\bibfnamefont{E.}~\bibnamefont{Perfetto}},
  \bibinfo{author}{\bibfnamefont{A.-M.} \bibnamefont{Uimonen}},
  \bibinfo{author}{\bibfnamefont{R.}~\bibnamefont{van Leeuwen}},
  \bibnamefont{and}
  \bibinfo{author}{\bibfnamefont{G.}~\bibnamefont{Stefanucci}},
  \bibinfo{journal}{Phys. Rev. B} \textbf{\bibinfo{volume}{89}},
  \bibinfo{pages}{075306} (\bibinfo{year}{2014}).

\bibitem[{\citenamefont{Perfetto et~al.}(2015)\citenamefont{Perfetto, Uimonen,
  van Leeuwen, and Stefanucci}}]{PUvLS.2015}
\bibinfo{author}{\bibfnamefont{E.}~\bibnamefont{Perfetto}},
  \bibinfo{author}{\bibfnamefont{A.-M.} \bibnamefont{Uimonen}},
  \bibinfo{author}{\bibfnamefont{R.}~\bibnamefont{van Leeuwen}},
  \bibnamefont{and}
  \bibinfo{author}{\bibfnamefont{G.}~\bibnamefont{Stefanucci}},
  \bibinfo{journal}{Phys. Rev. A} \textbf{\bibinfo{volume}{92}},
  \bibinfo{pages}{033419} (\bibinfo{year}{2015}),
  \urlprefix\url{https://link.aps.org/doi/10.1103/PhysRevA.92.033419}.

\bibitem[{\citenamefont{Bostr\"om et~al.}(2018)\citenamefont{Bostr\"om,
  Mikkelsen, Verdozzi, Perfetto, and Stefanucci}}]{C60paper2018}
\bibinfo{author}{\bibfnamefont{E.~V.} \bibnamefont{Bostr\"om}},
  \bibinfo{author}{\bibfnamefont{A.}~\bibnamefont{Mikkelsen}},
  \bibinfo{author}{\bibfnamefont{C.}~\bibnamefont{Verdozzi}},
  \bibinfo{author}{\bibfnamefont{E.}~\bibnamefont{Perfetto}}, \bibnamefont{and}
  \bibinfo{author}{\bibfnamefont{G.}~\bibnamefont{Stefanucci}},
  \bibinfo{journal}{Nano Lett.} \textbf{\bibinfo{volume}{18}},
  \bibinfo{pages}{785} (\bibinfo{year}{2018}),
  \eprint{https://doi.org/10.1021/acs.nanolett.7b03995},
  \urlprefix\url{https://doi.org/10.1021/acs.nanolett.7b03995}.

\bibitem[{\citenamefont{Perfetto et~al.}(2018)\citenamefont{Perfetto, Sangalli,
  Marini, and Stefanucci}}]{PSMP.2018}
\bibinfo{author}{\bibfnamefont{E.}~\bibnamefont{Perfetto}},
  \bibinfo{author}{\bibfnamefont{D.}~\bibnamefont{Sangalli}},
  \bibinfo{author}{\bibfnamefont{A.}~\bibnamefont{Marini}}, \bibnamefont{and}
  \bibinfo{author}{\bibfnamefont{G.}~\bibnamefont{Stefanucci}},
  \bibinfo{journal}{The Journal of Physical Chemistry Letters}
  \textbf{\bibinfo{volume}{9}}, \bibinfo{pages}{1353} (\bibinfo{year}{2018}),
  \eprint{https://doi.org/10.1021/acs.jpclett.8b00025},
  \urlprefix\url{https://doi.org/10.1021/acs.jpclett.8b00025}.

\bibitem[{\citenamefont{Lipavsk\'y et~al.}(1986)\citenamefont{Lipavsk\'y,
  \ifmmode \check{S}\else \v{S}\fi{}pi\ifmmode~\check{c}\else \v{c}\fi{}ka, and
  Velick\'y}}]{PhysRevB.34.6933}
\bibinfo{author}{\bibfnamefont{P.}~\bibnamefont{Lipavsk\'y}},
  \bibinfo{author}{\bibfnamefont{V.}~\bibnamefont{\ifmmode \check{S}\else
  \v{S}\fi{}pi\ifmmode~\check{c}\else \v{c}\fi{}ka}}, \bibnamefont{and}
  \bibinfo{author}{\bibfnamefont{B.}~\bibnamefont{Velick\'y}},
  \bibinfo{journal}{Phys. Rev. B} \textbf{\bibinfo{volume}{34}},
  \bibinfo{pages}{6933} (\bibinfo{year}{1986}),
  \urlprefix\url{https://link.aps.org/doi/10.1103/PhysRevB.34.6933}.

\bibitem[{\citenamefont{Almbladh et~al.}(1989)\citenamefont{Almbladh, Morales,
  and Grossmann}}]{PhysRevB.39.3489}
\bibinfo{author}{\bibfnamefont{C.-O.} \bibnamefont{Almbladh}},
  \bibinfo{author}{\bibfnamefont{A.~L.} \bibnamefont{Morales}},
  \bibnamefont{and}
  \bibinfo{author}{\bibfnamefont{G.}~\bibnamefont{Grossmann}},
  \bibinfo{journal}{Phys. Rev. B} \textbf{\bibinfo{volume}{39}},
  \bibinfo{pages}{3489} (\bibinfo{year}{1989}),
  \urlprefix\url{https://link.aps.org/doi/10.1103/PhysRevB.39.3489}.

\bibitem[{\citenamefont{Almbladh and Morales}(1989)}]{PhysRevB.39.3503}
\bibinfo{author}{\bibfnamefont{C.-O.} \bibnamefont{Almbladh}} \bibnamefont{and}
  \bibinfo{author}{\bibfnamefont{A.~L.} \bibnamefont{Morales}},
  \bibinfo{journal}{Phys. Rev. B} \textbf{\bibinfo{volume}{39}},
  \bibinfo{pages}{3503} (\bibinfo{year}{1989}),
  \urlprefix\url{https://link.aps.org/doi/10.1103/PhysRevB.39.3503}.

\bibitem[{\citenamefont{Cini}(1993)}]{cini1993two}
\bibinfo{author}{\bibfnamefont{M.}~\bibnamefont{Cini}}, \bibinfo{journal}{Solid
  state communications} \textbf{\bibinfo{volume}{88}}, \bibinfo{pages}{1101}
  (\bibinfo{year}{1993}).

\bibitem[{\citenamefont{Sawatzky}(1977)}]{PhysRevLett.39.504}
\bibinfo{author}{\bibfnamefont{G.~A.} \bibnamefont{Sawatzky}},
  \bibinfo{journal}{Phys. Rev. Lett.} \textbf{\bibinfo{volume}{39}},
  \bibinfo{pages}{504} (\bibinfo{year}{1977}),
  \urlprefix\url{https://link.aps.org/doi/10.1103/PhysRevLett.39.504}.

\bibitem[{\citenamefont{Balzer et~al.}(2010)\citenamefont{Balzer, Bauch, and
  Bonitz}}]{PhysRevA.82.033427}
\bibinfo{author}{\bibfnamefont{K.}~\bibnamefont{Balzer}},
  \bibinfo{author}{\bibfnamefont{S.}~\bibnamefont{Bauch}}, \bibnamefont{and}
  \bibinfo{author}{\bibfnamefont{M.}~\bibnamefont{Bonitz}},
  \bibinfo{journal}{Phys. Rev. A} \textbf{\bibinfo{volume}{82}},
  \bibinfo{pages}{033427} (\bibinfo{year}{2010}),
  \urlprefix\url{https://link.aps.org/doi/10.1103/PhysRevA.82.033427}.

\bibitem[{\citenamefont{Balzer et~al.}(2012)\citenamefont{Balzer, Hermanns, and
  Bonitz}}]{BalzerHermanns2012}
\bibinfo{author}{\bibfnamefont{K.}~\bibnamefont{Balzer}},
  \bibinfo{author}{\bibfnamefont{S.}~\bibnamefont{Hermanns}}, \bibnamefont{and}
  \bibinfo{author}{\bibfnamefont{M.}~\bibnamefont{Bonitz}},
  \bibinfo{journal}{EPL (Europhysics Letters)} \textbf{\bibinfo{volume}{98}},
  \bibinfo{pages}{67002} (\bibinfo{year}{2012}),
  \urlprefix\url{http://stacks.iop.org/0295-5075/98/i=6/a=67002}.

\bibitem[{\citenamefont{S\"akkinen et~al.}(2012)\citenamefont{S\"akkinen,
  Manninen, and van Leeuwen}}]{Sakkinen-2012}
\bibinfo{author}{\bibfnamefont{N.}~\bibnamefont{S\"akkinen}},
  \bibinfo{author}{\bibfnamefont{M.}~\bibnamefont{Manninen}}, \bibnamefont{and}
  \bibinfo{author}{\bibfnamefont{R.}~\bibnamefont{van Leeuwen}},
  \bibinfo{journal}{New Journal of Physics} \textbf{\bibinfo{volume}{14}},
  \bibinfo{pages}{013032} (\bibinfo{year}{2012}),
  \urlprefix\url{http://stacks.iop.org/1367-2630/14/i=1/a=013032}.

\bibitem[{\citenamefont{Hermanns et~al.}(2014)\citenamefont{Hermanns,
  Schl\"unzen, and Bonitz}}]{HermannsPRB2014}
\bibinfo{author}{\bibfnamefont{S.}~\bibnamefont{Hermanns}},
  \bibinfo{author}{\bibfnamefont{N.}~\bibnamefont{Schl\"unzen}},
  \bibnamefont{and} \bibinfo{author}{\bibfnamefont{M.}~\bibnamefont{Bonitz}},
  \bibinfo{journal}{Phys. Rev. B} \textbf{\bibinfo{volume}{90}},
  \bibinfo{pages}{125111} (\bibinfo{year}{2014}),
  \urlprefix\url{https://link.aps.org/doi/10.1103/PhysRevB.90.125111}.

\bibitem[{\citenamefont{Schl\"unzen and Bonitz}(2016)}]{CTPPBonitz2016}
\bibinfo{author}{\bibfnamefont{N.}~\bibnamefont{Schl\"unzen}} \bibnamefont{and}
  \bibinfo{author}{\bibfnamefont{M.}~\bibnamefont{Bonitz}},
  \bibinfo{journal}{Contrib. Plasma Phys.} \textbf{\bibinfo{volume}{56}},
  \bibinfo{pages}{5} (\bibinfo{year}{2016}), ISSN \bibinfo{issn}{1521-3986},
  \urlprefix\url{http://dx.doi.org/10.1002/ctpp.201610003}.

\bibitem[{\citenamefont{Hopjan et~al.}(2016)\citenamefont{Hopjan, Karlsson,
  Ydman, Verdozzi, and Almbladh}}]{HopjanPRL2016}
\bibinfo{author}{\bibfnamefont{M.}~\bibnamefont{Hopjan}},
  \bibinfo{author}{\bibfnamefont{D.}~\bibnamefont{Karlsson}},
  \bibinfo{author}{\bibfnamefont{S.}~\bibnamefont{Ydman}},
  \bibinfo{author}{\bibfnamefont{C.}~\bibnamefont{Verdozzi}}, \bibnamefont{and}
  \bibinfo{author}{\bibfnamefont{C.-O.} \bibnamefont{Almbladh}},
  \bibinfo{journal}{Phys. Rev. Lett.} \textbf{\bibinfo{volume}{116}},
  \bibinfo{pages}{236402} (\bibinfo{year}{2016}),
  \urlprefix\url{https://link.aps.org/doi/10.1103/PhysRevLett.116.236402}.

\bibitem[{\citenamefont{Lev and Reichman}(2016)}]{ReichmanEPL2016}
\bibinfo{author}{\bibfnamefont{Y.~B.} \bibnamefont{Lev}} \bibnamefont{and}
  \bibinfo{author}{\bibfnamefont{D.~R.} \bibnamefont{Reichman}},
  \bibinfo{journal}{EPL (Europhysics Letters)} \textbf{\bibinfo{volume}{113}},
  \bibinfo{pages}{46001} (\bibinfo{year}{2016}),
  \urlprefix\url{http://stacks.iop.org/0295-5075/113/i=4/a=46001}.

\bibitem[{\citenamefont{Schl\"unzen et~al.}(2017)\citenamefont{Schl\"unzen,
  Joost, Heidrich-Meisner, and Bonitz}}]{Joost2017}
\bibinfo{author}{\bibfnamefont{N.}~\bibnamefont{Schl\"unzen}},
  \bibinfo{author}{\bibfnamefont{J.-P.} \bibnamefont{Joost}},
  \bibinfo{author}{\bibfnamefont{F.}~\bibnamefont{Heidrich-Meisner}},
  \bibnamefont{and} \bibinfo{author}{\bibfnamefont{M.}~\bibnamefont{Bonitz}},
  \bibinfo{journal}{Phys. Rev. B} \textbf{\bibinfo{volume}{95}},
  \bibinfo{pages}{165139} (\bibinfo{year}{2017}),
  \urlprefix\url{https://link.aps.org/doi/10.1103/PhysRevB.95.165139}.

\bibitem[{\citenamefont{{A.-M.~Uimonen}
  et~al.}(2011)\citenamefont{{A.-M.~Uimonen}, Khosravi, Stan, Stefanucci,
  Kurth, {R.~van~Leeuwen}, and {E.K.U.~Gross}}}]{UKSSKvLG.2011}
\bibinfo{author}{\bibnamefont{{A.-M.~Uimonen}}},
  \bibinfo{author}{\bibfnamefont{E.}~\bibnamefont{Khosravi}},
  \bibinfo{author}{\bibfnamefont{A.}~\bibnamefont{Stan}},
  \bibinfo{author}{\bibfnamefont{G.}~\bibnamefont{Stefanucci}},
  \bibinfo{author}{\bibfnamefont{S.}~\bibnamefont{Kurth}},
  \bibinfo{author}{\bibnamefont{{R.~van~Leeuwen}}}, \bibnamefont{and}
  \bibinfo{author}{\bibnamefont{{E.K.U.~Gross}}}, \bibinfo{journal}{Phys. Rev.
  B} \textbf{\bibinfo{volume}{84}}, \bibinfo{pages}{115103}
  (\bibinfo{year}{2011}).

\end{thebibliography}
\end{document}